\documentclass[conference]{IEEEtran}
\IEEEoverridecommandlockouts
\usepackage{cite}
\usepackage{float}
\usepackage[ruled,vlined]{algorithm2e}
\usepackage{amsmath,amssymb,amsfonts}
\usepackage{algorithmic}
\usepackage{graphicx}
\usepackage{textcomp}
\usepackage{xcolor}
\def\BibTeX{{\rm B\kern-.05em{\sc i\kern-.025em b}\kern-.08em
    T\kern-.1667em\lower.7ex\hbox{E}\kern-.125emX}}
\begin{document}
\DeclareGraphicsExtensions{.pdf,.png,.jpg}
\setkeys{Gin}{width=\columnwidth} 

\title{Long-Term Spatio-Temporal Forecasting of Monthly Rainfall in West Bengal Using Ensemble Learning Approaches\\

}

\author{\IEEEauthorblockN{1\textsuperscript{st} Jishu Adhikary}
\IEEEauthorblockA{\textit{Financial Modelling Quantative Analyst} \\
\textit{UBS (Union Bank of Switzerland)}\\
Kolkata, India \\
jishuadhikary@gmail.com}
\and
\IEEEauthorblockN{2\textsuperscript{nd}  Dr. Raju Maiti}
\IEEEauthorblockA{\textit{Economic Research Unit} \\
\textit{Indian Statistical Institute}\\
Kolkata, India \\
rajumaiti@gmail.com}

}

\maketitle

\begin{abstract}
Rainfall forecasting plays a critical role in climate adaptation, agriculture, and water resource management. This study develops long-term forecasts of monthly rainfall across 19 districts of West Bengal using a century-scale dataset spanning 1900–2019. Daily rainfall records are aggregated into monthly series, resulting in 120 years of observations for each district. The forecasting task involves predicting the next 108 months (9 years, 2011-2019) while accounting for temporal dependencies and spatial interactions among districts. To address the nonlinear and complex structure of rainfall dynamics, we propose a hierarchical modeling framework that combines regression-based forecasting of yearly features with multi-layer perceptrons (MLPs) for monthly prediction. Yearly features—such as annual totals, quarterly proportions, variability measures, skewness, and extremes—are first forecasted using regression models that incorporate both own lags and neighboring-district lags. These forecasts are then integrated as auxiliary inputs into an MLP model, which captures nonlinear temporal patterns and spatial dependencies in the monthly series. The results demonstrate that the hierarchical regression–MLP architecture provides robust long-term spatio-temporal forecasts, offering valuable insights for agriculture, irrigation planning, and water conservation strategies.

\end{abstract}

\begin{IEEEkeywords}
Rainfall forecasting, Spatio-temporal modeling, Hierarchical neural networks, MLP, Long-term prediction, Climate variability, West Bengal.
\end{IEEEkeywords}

\section{Introduction}
Rainfall is one of the most critical climatic variables influencing agriculture, water resources, and socio-economic systems, particularly in regions such as West Bengal, India, where livelihoods depend heavily on the monsoon. Reliable rainfall forecasting supports timely decisions for irrigation planning, crop management, flood preparedness, and disaster risk reduction. While short-term rainfall prediction has been widely studied, long-term forecasting at regional and district levels remains a challenging task due to strong variability, seasonality, and spatial heterogeneity in rainfall behavior.

This study develops a spatio-temporal framework for long-term rainfall forecasting across 19 districts of West Bengal. The dataset consists of daily rainfall records spanning 1900–2019, which are aggregated into monthly series to provide 120 years of observations for each district. The forecasting horizon extends over the next 108 months (9 years, 2011-2019), requiring models that can simultaneously capture complex temporal dependencies and spatial correlations among districts. Neighboring districts often display related rainfall patterns due to shared climatic and geographical factors, making it important to explicitly incorporate inter-district relationships in the modeling design.

The proposed framework is based on a hierarchical modeling architecture that combines regression methods with multi-layer perceptrons (MLPs). In the first stage, yearly features are derived from the monthly rainfall series, producing secondary time series that describe broader aspects of annual rainfall behavior. These features include yearly totals, quarterly proportions, measures of variability such as standard deviation and entropy, and extremes such as annual maxima. Each yearly feature series is smoothed and forecasted for individual districts using regression models that rely on both own lags and lagged values from neighboring districts. In the second stage, the resulting forecasts of yearly features serve as auxiliary inputs to an MLP model, which generates monthly rainfall predictions. This design enables the model to integrate nonlinear temporal dynamics with higher-level annual characteristics, improving robustness in long-term forecasting.

To evaluate performance, the proposed approach is compared against two challenger models: (i) baseline model -  a naïve seasonal model  fitted separately to each district’s series, and (ii) benchmark model - a standard MLP model that uses lagged values of each district and its neighbors without yearly feature integration. Comparative results demonstrate that the proposed hierarchical regression–MLP architecture more effectively captures both temporal evolution and spatial dependencies in rainfall, leading to improved long-term forecasts. 

The remainder of this paper is organized as follows: Section II reviews related work on rainfall  forecasting and spatio-temporal modeling. Section III describes the dataset and the preprocessing procedures. Section IV presents the proposed hierarchical regression–MLP framework, including the construction of yearly features and the forecasting architecture. Section V reports the experimental setup, results, and comparative evaluation with baseline models. Finally, Section VI concludes the study and highlights directions for future research.

\section{Related Work}

Rainfall forecasting has long been a topic of significant research interest owing to its relevance for agriculture, water management, flood preparedness, and climate adaptation. Over the years, approaches have evolved from classical statistical models to modern machine learning and deep learning architectures, with recent emphasis on spatio-temporal and physics-informed frameworks. This section reviews key contributions across these domains.

\subsection{Statistical and Classical Models}

Early rainfall forecasting relied heavily on statistical time series models such as ARIMA and SARIMA \cite{b1}. These methods capture linear temporal dependencies but struggle with nonlinearities and long-term variability. Multivariate methods, such as Vector Autoregression (VAR), have been applied to study spatial linkages between stations \cite{b4}. Hybrid statistical–neural approaches, like ARIMA–LSTM \cite{b8}, have been proposed to improve long-term forecasts by combining trend-based and nonlinear components. Downscaling methods, such as DeepSD and CNN-based statistical downscalers, have also been explored for high-resolution Indian monsoon rainfall \cite{b5}.

\subsection{Machine Learning Approaches}

Machine learning methods, including Random Forest (RF), Support Vector Regression (SVR), and Gradient Boosting models, have been extensively employed to capture nonlinear rainfall behavior \cite{b5,b10}. Comparative studies indicate that machine learning models outperform classical methods in short-term and regional rainfall prediction \cite{b10}. Ensemble and hybrid ML frameworks integrating Random Forests with physical predictors have demonstrated robustness in district-level forecasts. Probabilistic ensemble forecasting frameworks have also been applied for uncertainty quantification \cite{b14}.

\subsection{Deep Learning Models}

Deep learning methods have gained prominence for rainfall prediction due to their ability to extract high-level temporal and spatial features \cite{b11}. Recurrent Neural Networks (RNNs) and Long Short-Term Memory (LSTM) architectures \cite{b2} have shown superiority in capturing long-term dependencies in rainfall time series. Spatial-temporal attention-based LSTM and ConvLSTM networks further enhance predictive accuracy by integrating spatial information \cite{b3,b12}. CNN–LSTM hybrids have been applied to Indian monsoon prediction, effectively combining spatial and temporal components \cite{b12}. However, these architectures often require extensive data and computation, making them challenging for century-scale modeling.

\subsection{Spatio-Temporal and Graph-Based Frameworks}

Recent work increasingly emphasizes the joint modeling of spatial and temporal rainfall patterns. Graph Neural Networks (GNNs) have been successfully used for distributed hydrological forecasting \cite{b9} and for coupling physical and spatial factors in precipitation modeling \cite{b15}. Hierarchical spatio-temporal GNNs (HiSTGNN) further improve forecasting by incorporating multi-level spatial hierarchies \cite{b16}. Multi-modal GNN frameworks that combine satellite, station, and reanalysis data have been shown to enhance off-grid rainfall predictions \cite{b18}. Ensemble post-processing with GNNs has also been explored for improving extreme rainfall forecasting \cite{b17}.

\subsection{Hybrid and Physics-Informed Models}

Hybrid and physics-informed models integrate physical constraints with machine learning to enhance interpretability and generalization. Physics-guided GANs and coupled $\omega$-GNN frameworks have demonstrated improved robustness in simulating extreme precipitation events \cite{b15,b17}. Probabilistic and Bayesian deep learning models have also been used for quantifying uncertainty in rainfall prediction \cite{b14}. These hybrid strategies represent a promising direction for district-scale, long-term hydrological modeling.

\subsection{Research Gap}

Despite these advances, challenges persist in long-term, district-level spatio-temporal forecasting. Most prior studies focus either on high-frequency (daily) or coarse-resolution (regional) scales. Limited attention has been given to integrating higher-level yearly rainfall characteristics with monthly temporal patterns in a unified structure. Furthermore, applications using century-scale datasets remain scarce, particularly those capturing evolving spatial dependencies and climatic non-stationarities. To bridge these gaps, the present study proposes a hierarchical modeling approach that couples regression-based yearly feature prediction with multilayer perceptrons (MLPs) for monthly forecasting. This allows nonlinear temporal patterns and spatial correlations among districts to be captured more effectively within a unified spatio-temporal framework.

\section{Data Description }

\subsection{Data Source and Study Area}

West Bengal, situated in eastern India, experiences a monsoon-dominated climate with marked spatial and seasonal variability in rainfall. Coastal districts such as Purba Medinipur and North 24 Parganas receive higher rainfall due to their proximity to the Bay of Bengal, while western districts like Purulia and Bankura are comparatively drier. In North Bengal, districts including Darjeeling and Jalpaiguri receive higher precipitation from both the southwest (June–September) and northeast (October–November) monsoons, with orographic effects further enhancing rainfall.

\begin{table}[h]
\centering
\caption{Number of Stations in Each District}
\begin{tabular}{l c}
\hline
\textbf{District} & \textbf{No. of Stations} \\
\hline
Bankura          & 31 \\
Darjeeling       & 30 \\
Jalpaiguri       & 29 \\
West Midnapore   & 26 \\
East Midnapore   & 21 \\
Purulia  & 19 \\
Murshidabad      & 18 \\
Burdwan          & 18 \\
Birbhum          & 16 \\
Hooghly          & 14 \\
Cooch Behar      & 11 \\
24 Parganas S    & 11 \\
24 Parganas N    & 10 \\
Dinajpur North   &  9 \\
Dinajpur South   &  9 \\
Malda            &  8 \\
Nadia            &  7 \\
Howrah           &  5 \\
Kolkata          &  1 \\
\hline
\end{tabular}
\label{tab:dist_wise_stations}
\end{table}

The dataset comprises daily rainfall observations spanning 1900–2019, recorded at 294 stations distributed across 19 districts.  Table~\ref{tab:dist_wise_stations} lists the number of stations per district, ranging from 31 in Bankura to a single station in Kolkata. As daily rainfall data are sparse, with many zero-rainfall entries, the series are aggregated to a monthly scale, yielding 1440 time points (120 years × 12 months) per district. This results in 19 district-level time series that capture long-term temporal dynamics while retaining spatial heterogeneity.

These district-level series exhibit both temporal trends and spatial correlations, with neighboring districts often showing similar rainfall evolution. Such dependencies motivate the need for spatio-temporal forecasting approaches.

\subsection{Data Aggregation and Transformation}
The raw dataset consists of daily rainfall observations recorded at 294 meteorological stations distributed across 19 districts of West Bengal. Since multiple observatories exist within a single district, daily records often exhibit heterogeneity; for example, rainfall may be reported in one station while another station in the same district records no precipitation. To obtain a consistent district-level representation, we first aggregated the station-level daily measurements into district-level daily rainfall values. For this purpose, the total daily rainfall across all stations within a district was considered, as it better reflects the overall water received in the region compared to a simple average, which may understate localized heavy rainfall events.

Once district-level daily rainfall series were constructed, the data were temporally aggregated to obtain monthly totals for each district. This two-step procedure—spatial aggregation from stations to districts, followed by temporal aggregation from days to months—effectively reduces the sparsity inherent in the raw dataset, where numerous days report zero rainfall or missing values. The resulting dataset thus provides 120 years (1900–2019) of monthly rainfall (in milimeter) time series  for 19 districts, forming the basis for subsequent forecasting analysis.

\begin{figure}[htbp]
    \centering
    \begin{minipage}{0.48\linewidth}
        \centering
        \includegraphics[width=\linewidth]{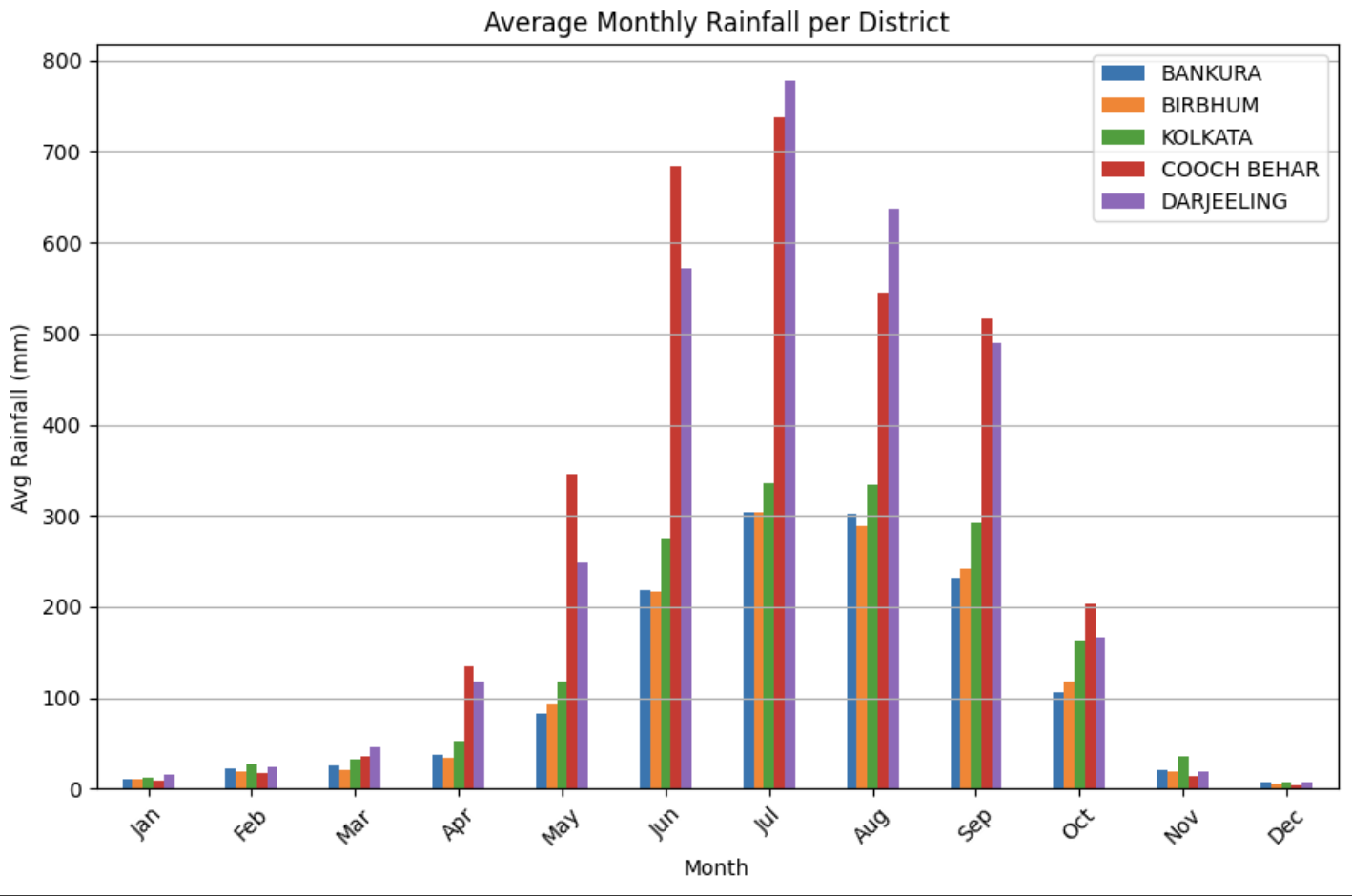}
        \caption{Average monthly rainfall over four decades for selected districts, showing a clear peak during the monsoon months (June–September).}
        \label{fig:barplot}
    \end{minipage}
    \hfill
    \begin{minipage}{0.45\linewidth}
        \centering
        \includegraphics[width=\linewidth]{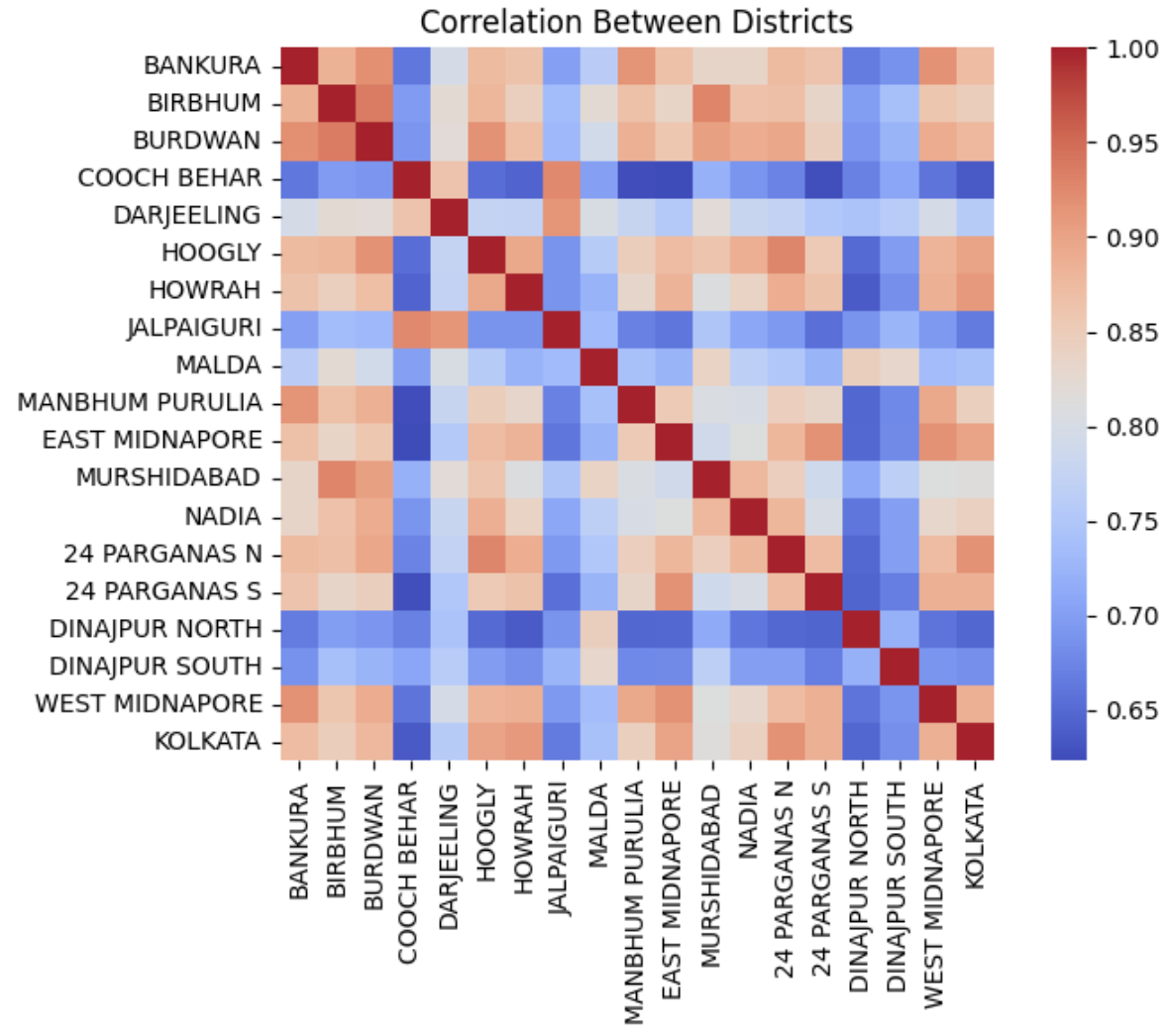}
        \caption{Correlation of monthly rainfall across districts, with stronger similarity among geographically close regions.}
        \label{fig:cor_mat}
    \end{minipage}
\end{figure}

\begin{figure}
    \centering
    \includegraphics[width=1\linewidth]{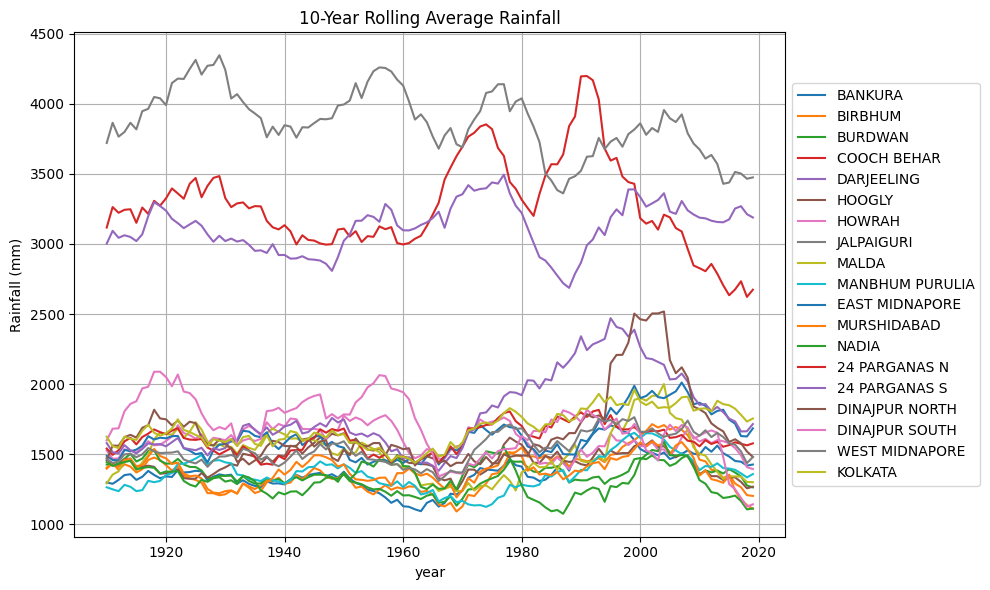}
    \caption{Ten-year moving average of annual rainfall across districts. Northern districts such as Darjeeling and Cooch Behar record higher rainfall with stronger fluctuations, while western districts remain lower and relatively stable.}
    \label{fig:ts_plot_1}
\end{figure}

\subsection{Data Partitioning}

For modeling and evaluation purposes, the dataset was divided into two distinct periods. 
The years 1900-2010 were designated as the \textit{training set}, which was used for exploratory 
data analysis, feature construction, and model calibration. The subsequent years, 2011-2019, 
were reserved as the \textit{holdout set} and were used exclusively for out-of-sample evaluation. 
This separation ensures that no information from the holdout period influences the training 
process or exploratory procedures, thereby providing an unbiased assessment of forecasting 
performance.

\section{Exploratory Data Analysis}

\subsection{Temporal Rainfall Patterns}

Rainfall in West Bengal shows a strong monsoon cycle, with most rain falling between June and September (see Fig.~\ref{fig:barplot}). Northern districts such as Cooch Behar and Darjeeling record the highest amounts, while western districts like Bankura and Birbhum remain much drier. The correlation plot (Fig.~\ref{fig:cor_mat}) confirms that nearby districts share similar rainfall patterns, especially in North Bengal, whereas the western plateau shows weaker connections.


\begin{figure}[h!]
    \centering
    \begin{tabular}{cc}
        \includegraphics[width=0.45\linewidth]{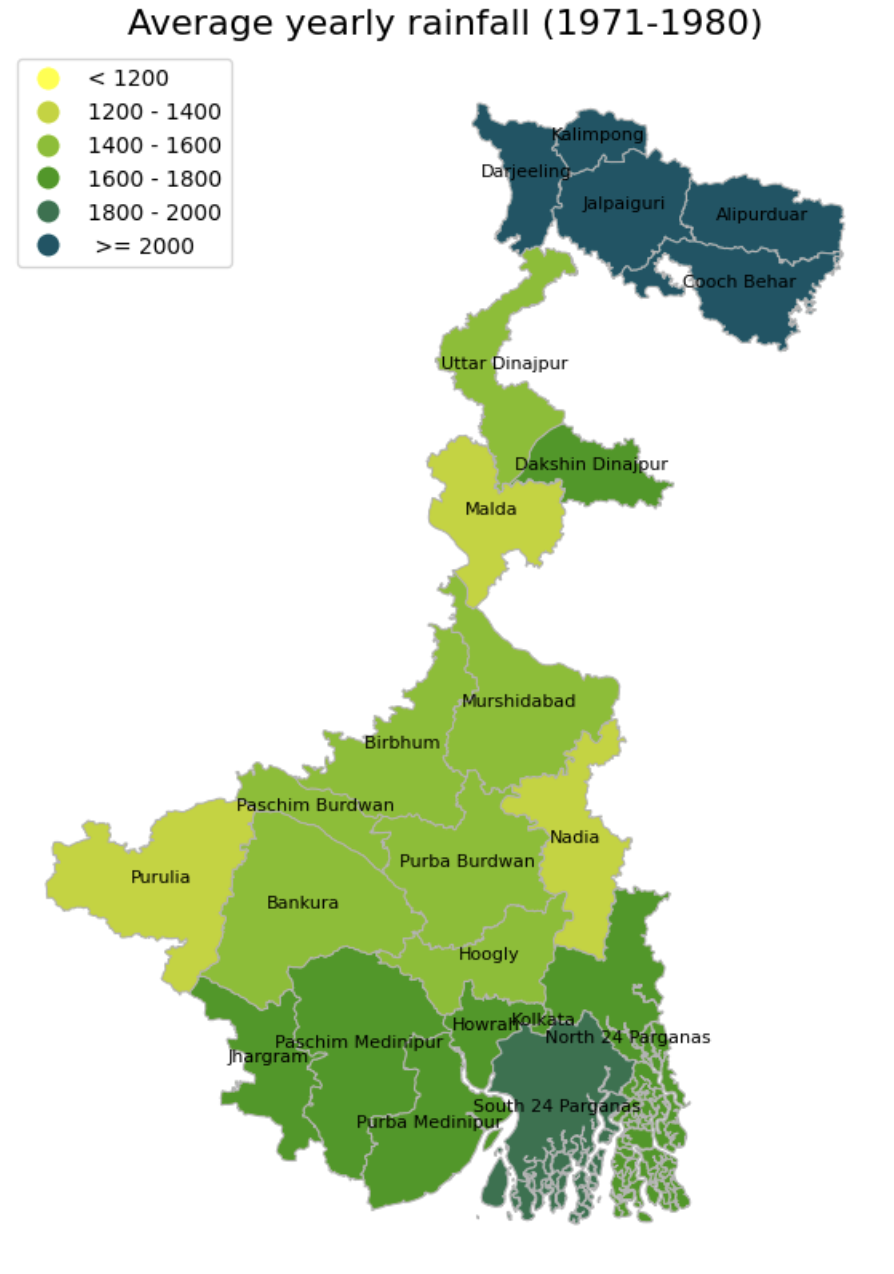} &
        \includegraphics[width=0.45\linewidth]{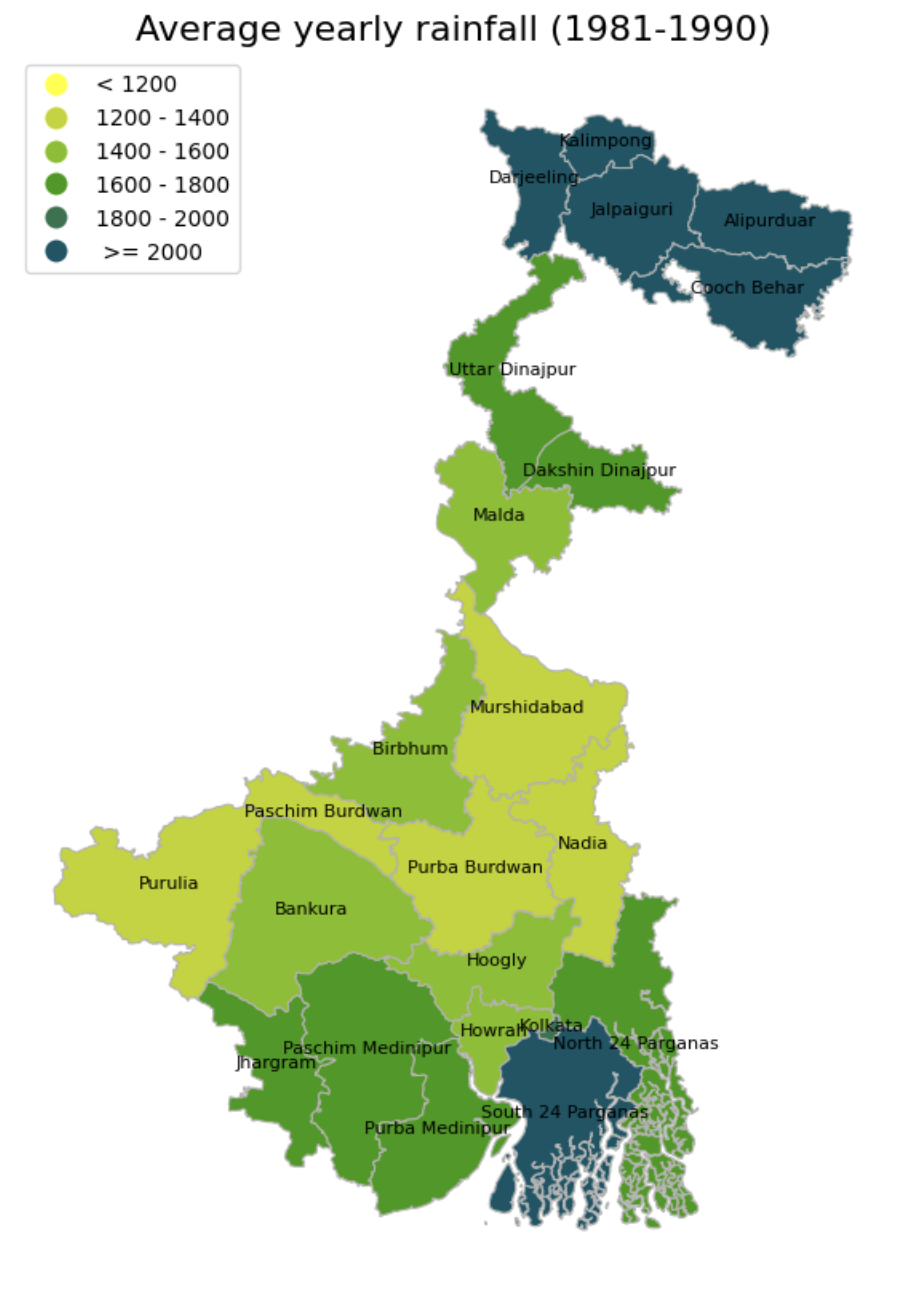} \\
        \includegraphics[width=0.45\linewidth]{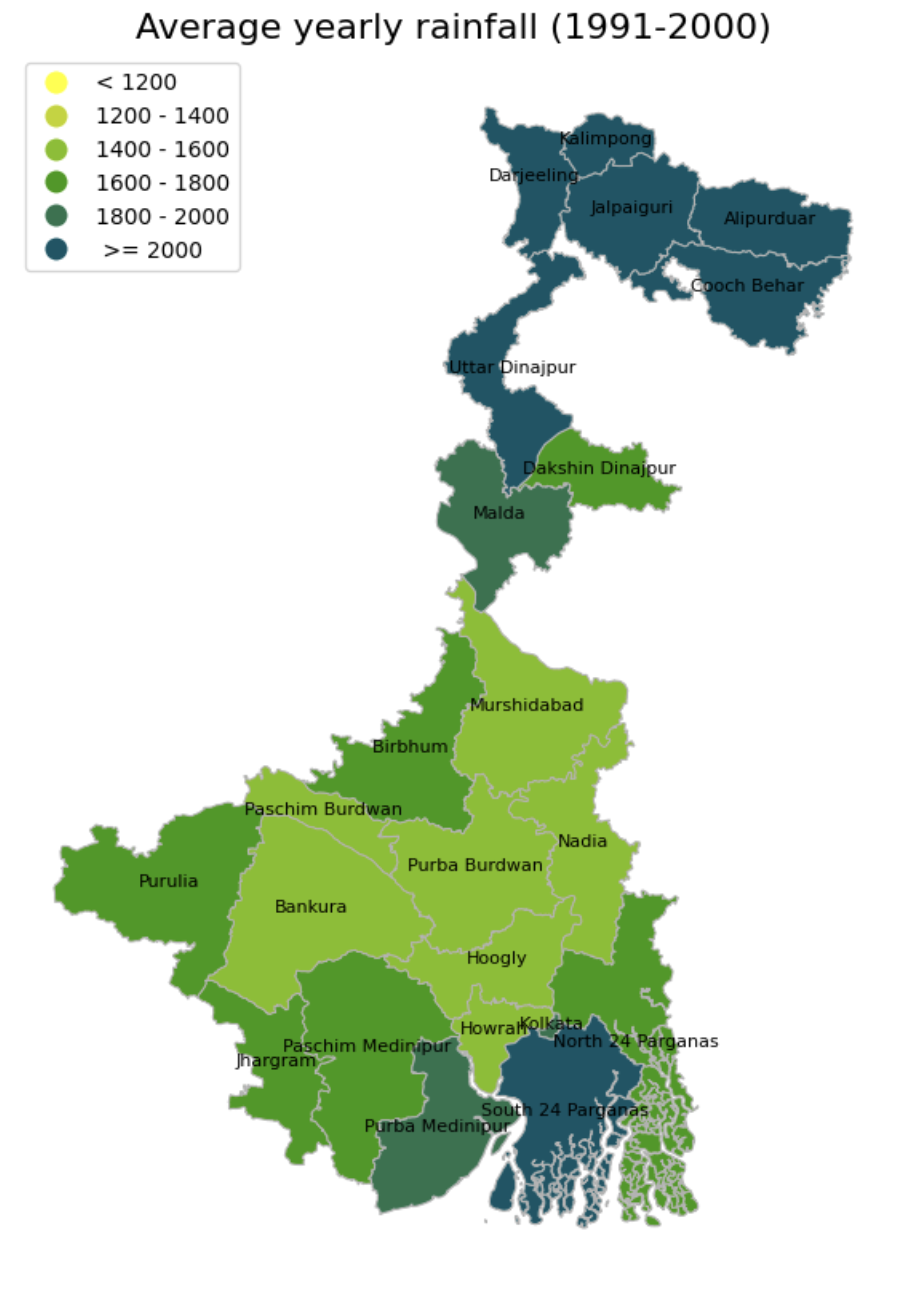} &
        \includegraphics[width=0.45\linewidth]{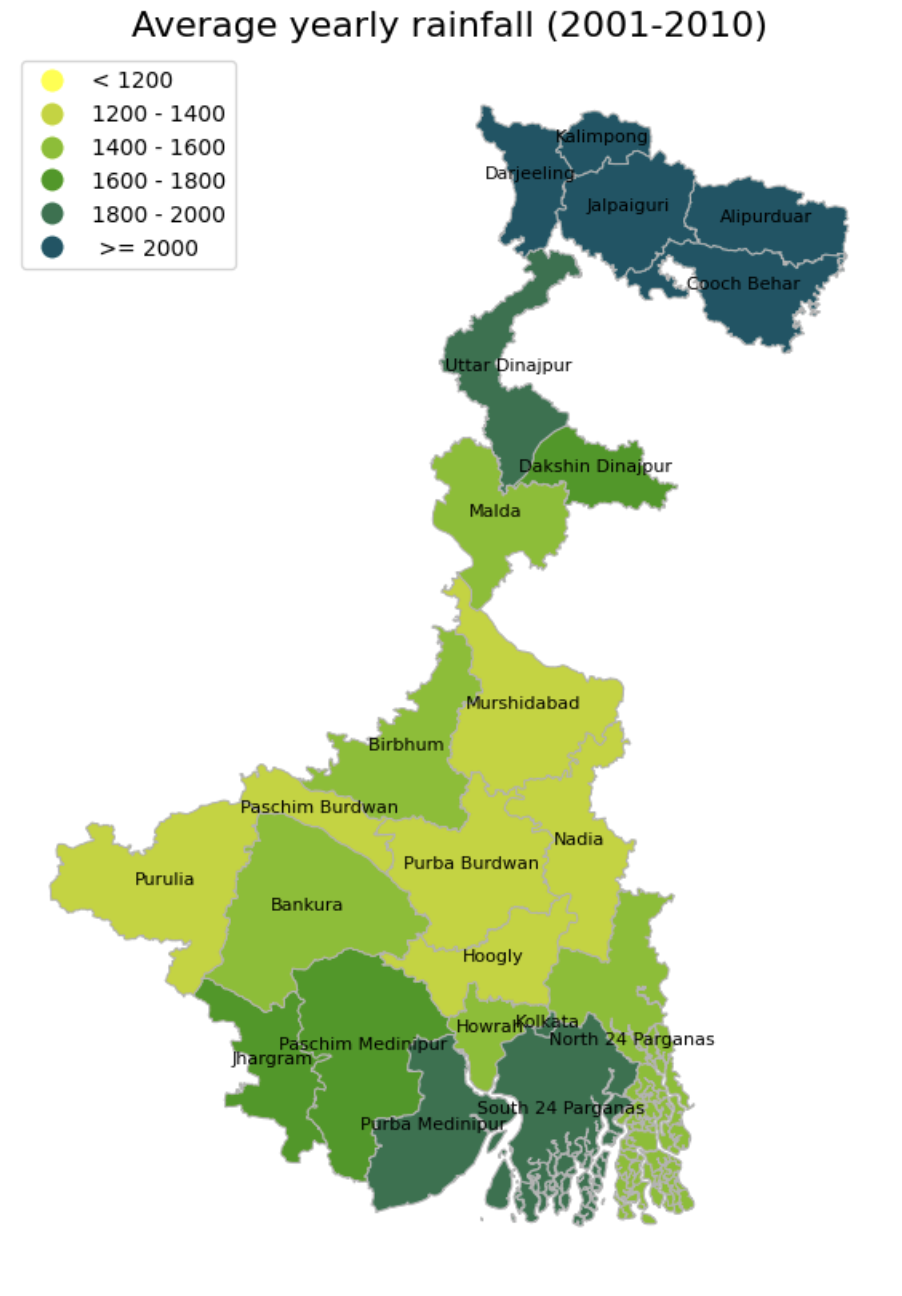} \\
    \end{tabular}
    \caption{Spatial distribution of average annual rainfall across West Bengal over four decades (1971–2010), shown separately for 1971–1980, 1981–1990, 1991–2000, and 2001–2010.}
    \label{fig:map_avg_yearly_rainfall}
\end{figure}


Long-term averages (see Fig.~\ref{fig:ts_plot_1}) highlight clear contrasts: Darjeeling and Cooch Behar receive over 3000–4000 mm annually, though both show large swings across decades, including a peak in the 1990s followed by decline. In comparison, western districts remain consistently low at 1200–1600 mm. Many districts show a gradual decrease after the 2000s, hinting at weakening monsoon intensity.

\textbf{}

\subsection{Spatio-Temporal Rainfall Dynamics: Decadal Analysis}

An in-depth analysis of rainfall patterns across different decades reveals significant spatio-temporal shifts crucial for understanding climate variability in the region.


\begin{figure}[h!]
    \centering
    \begin{tabular}{cc}
        \includegraphics[width=0.48\linewidth]{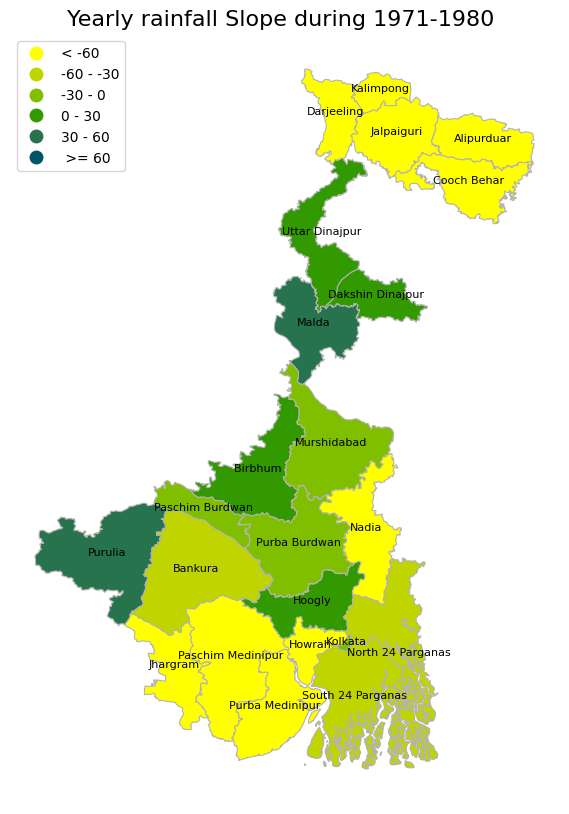} &
        \includegraphics[width=0.48\linewidth]{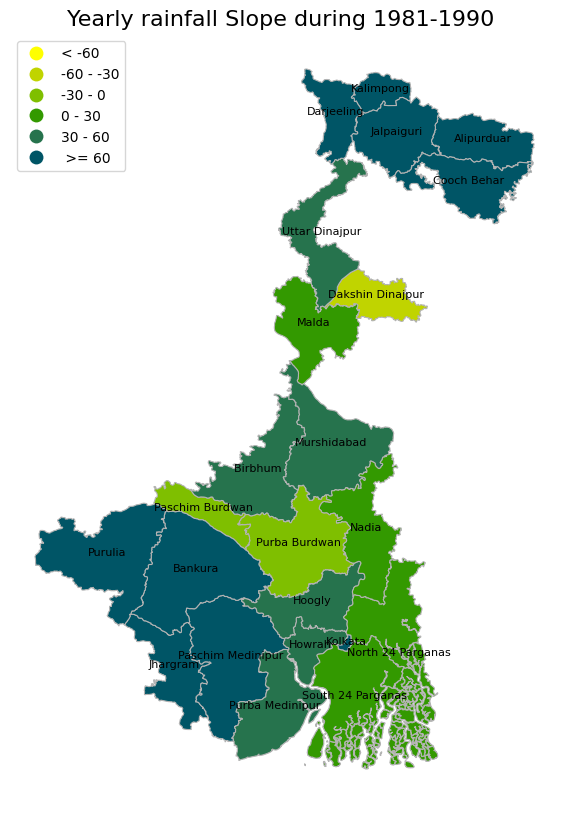} \\
        \includegraphics[width=0.48\linewidth]{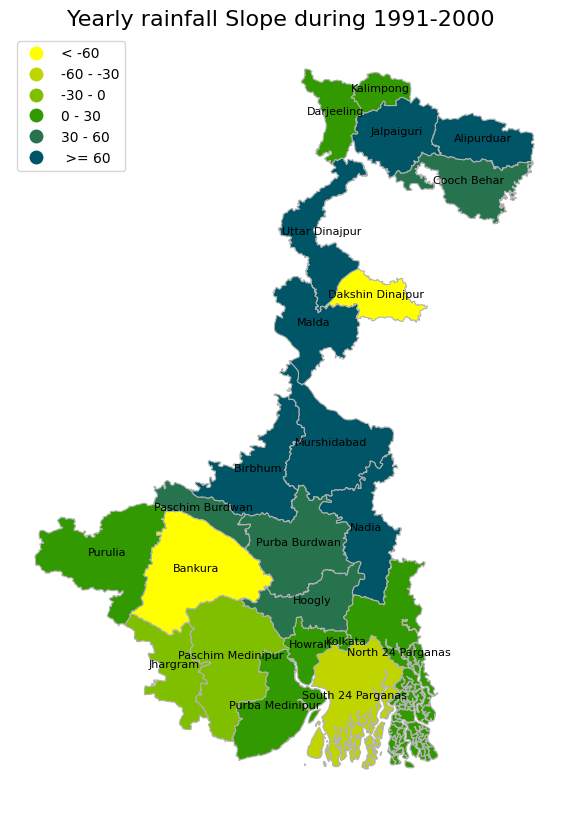} &
        \includegraphics[width=0.48\linewidth]{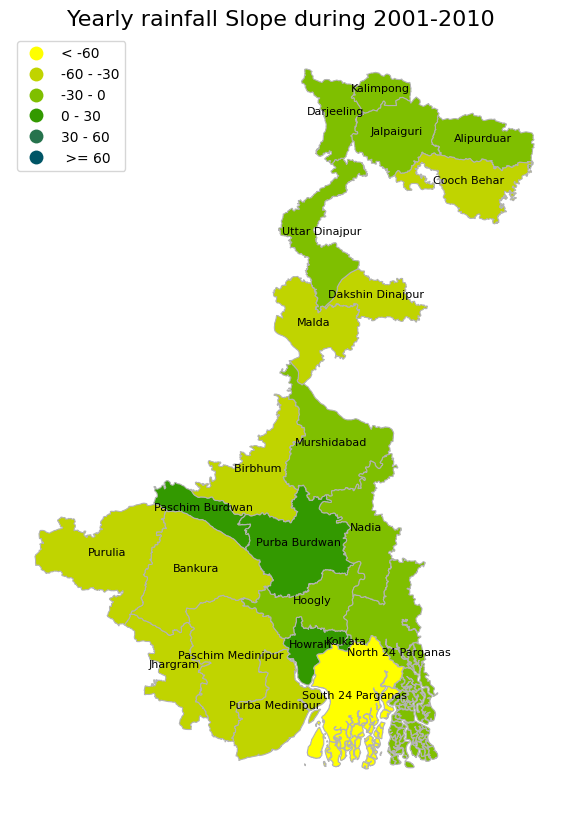} \\
    \end{tabular}
    \caption{Spatial distribution of slope of yearly rainfall across West Bengal
over four decades (1971–2010), shown separately for 1971–1980, 1981–1990,
1991–2000, and 2001–2010}
    \label{fig:map_avg_yearly_trend}
\end{figure}


\subsubsection{Trends in Overall Rainfall}

\setcounter{footnote}{0}

The maps in Fig.~\ref{fig:map_avg_yearly_rainfall}\footnote{The district boundaries in this map are based on the shapefile obtained from Kaggle. The shapefile reflects recent administrative updates in West Bengal, including the creation of Kalimpong (from Darjeeling), Jhargram (from Paschim Medinipur), and Alipurduar (from Jalpaiguri), as well as the division of Bardhaman into Purba (East) and Paschim (West) Bardhaman. However, the rainfall dataset provided by the West Bengal Pollution Control Board contains records for 19 districts based on the earlier administrative boundaries.}  showing average yearly rainfall across the decades highlight a  decline in precipitation intensity in southern and central Bengal. For instance, districts in the agriculturally significant Gangetic plains, such as  Bardhaman and Nadia, show a gradual shift to lower rainfall bands (1400–1600 mm) over time. This trend suggests a declining rainfall regime in these regions, which could significantly impact crop patterns and water management planning. In contrast, districts in the northern sub-Himalayan regions, like Jalpaiguri and Darjeeling, maintain consistently high rainfall, a stable trend likely due to their proximity to the Himalayas. The western districts of Bankura and Purulia also show a trend toward increasing dryness, highlighting their growing vulnerability to water scarcity. Overall, a broad trend of decreasing yearly rainfall is evident across the state, excluding the northern foothills.

\subsubsection{Rainfall Trend Analysis (Decadal Slope, 1971–2010)}

To quantify the long-term changes in rainfall, a trend analysis was performed for each district over four decades (1971–1980, 1981–1990, 1991–2000, and 2001–2010). For each decade and district, a linear regression model was fitted to the annual rainfall data, where the slope of the regression line reflects the magnitude and direction of the change. A positive slope ($b>0$) indicates an increasing trend, while a negative slope ($b<0$) indicates a decreasing trend. The color-coded maps (Fig.~\ref{fig:map_avg_yearly_trend}) of the decadal slope visualize these findings.

\begin{figure}[h!]
    \centering
    \begin{tabular}{cc}
        \includegraphics[width=0.48\linewidth]{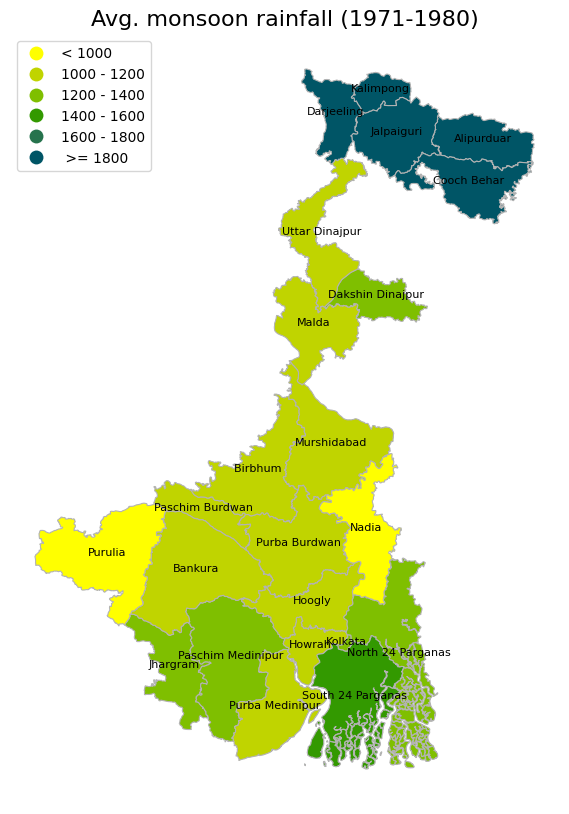} &
        \includegraphics[width=0.48\linewidth]{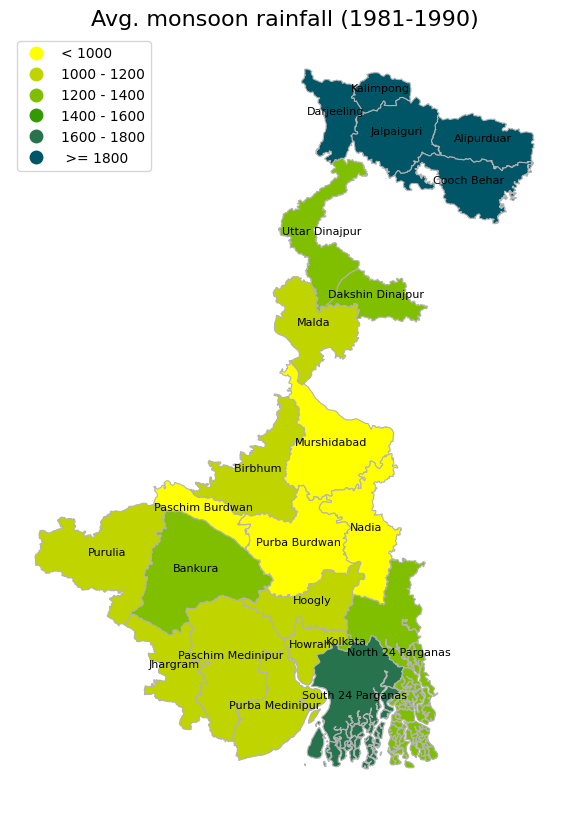} \\
        \includegraphics[width=0.48\linewidth]{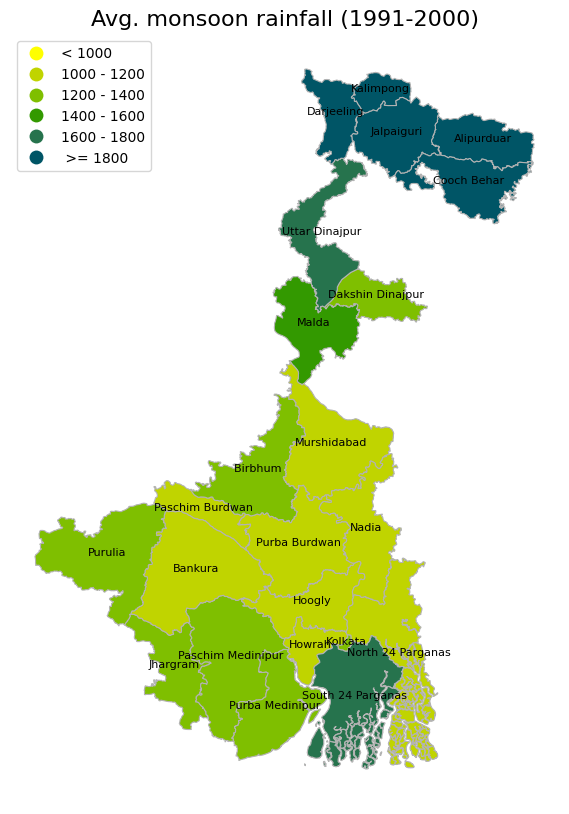} &
        \includegraphics[width=0.48\linewidth]{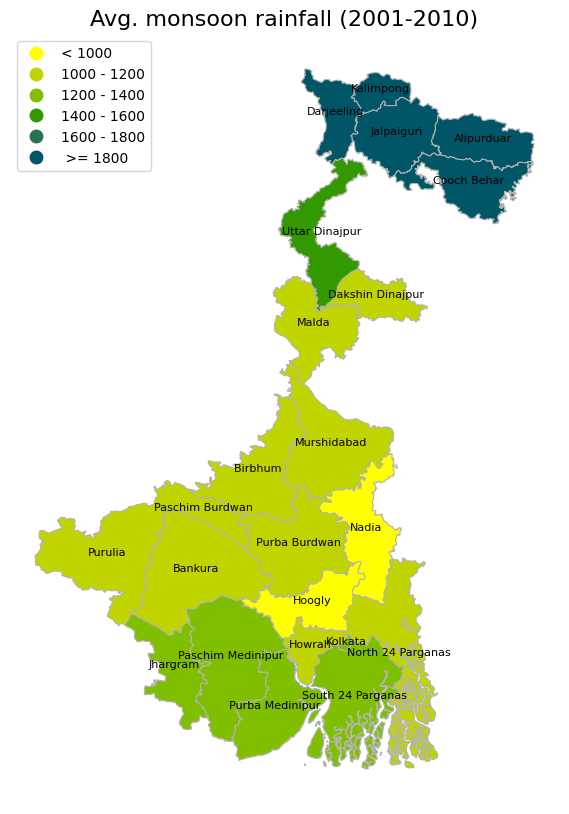} \\
    \end{tabular}
    \caption{Spatial distribution of average monsoon-season rainfall across West Bengal
over four decades (1971–2010), shown separately for 1971–1980, 1981–1990,
1991–2000, and 2001–2010}
    \label{fig:map_avg_yearly_monsoon}
\end{figure}

\textbf{Key Observations}
\begin{itemize}
    \item \textbf{1971–1980:} This period shows a mixed rainfall trend. While central districts like Burdwan and Midnapore exhibit a slight increasing trend (as shown by a positive slope), the southern and northern fringe districts (e.g., South 24 Parganas, Jalpaiguri) show declines.
    \item \textbf{1981–1990:} A widespread upward trend is evident during this decade, particularly in northern Bengal and southwestern districts like Darjeeling and Purulia. There are only a few patches of decreasing rainfall in the central areas.
    \item \textbf{1991–2000:} The upward trend from the previous decade continues, with a widespread increase in rainfall across the north and east. However, a notable reversal is observed in certain regions, with Bankura and South 24 Parganas showing an decreasing trend.
    \item \textbf{2001–2010:} A significant stabilization is observed, with the slopes for most districts nearing zero, indicating that the period of sharp incline has slowed. A  downward trend is evident in some southern and western districts, suggesting a potential decline from earlier upward trend.
\end{itemize}


\begin{figure}[h!]
    \centering
    \begin{tabular}{cc}
        \includegraphics[width=0.48\linewidth]{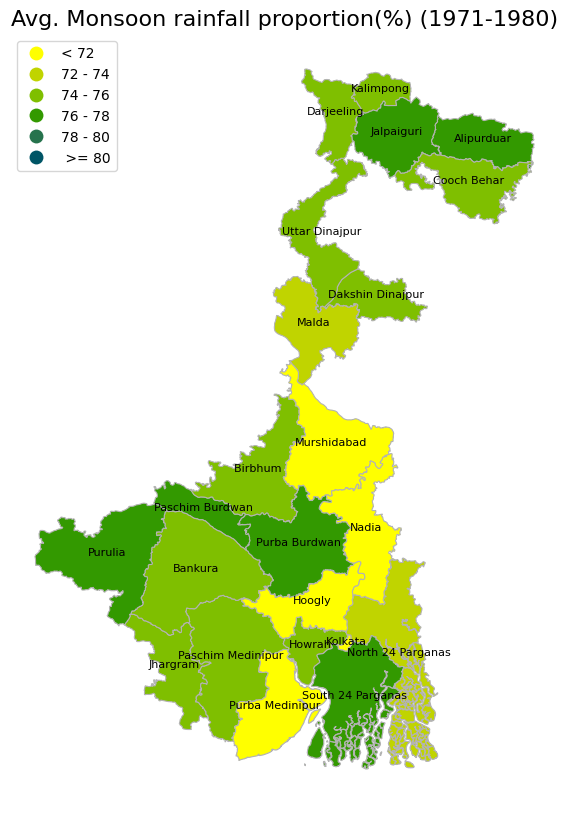} &
        \includegraphics[width=0.48\linewidth]{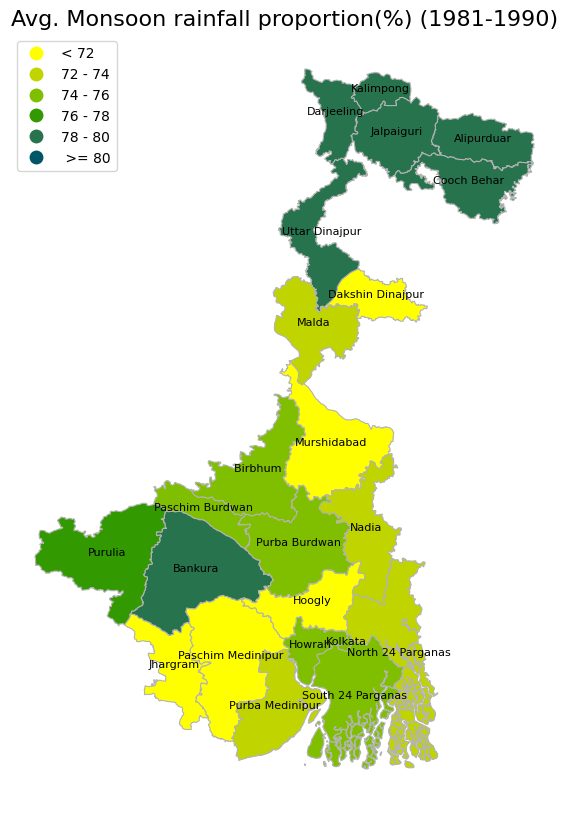} \\
        \includegraphics[width=0.48\linewidth]{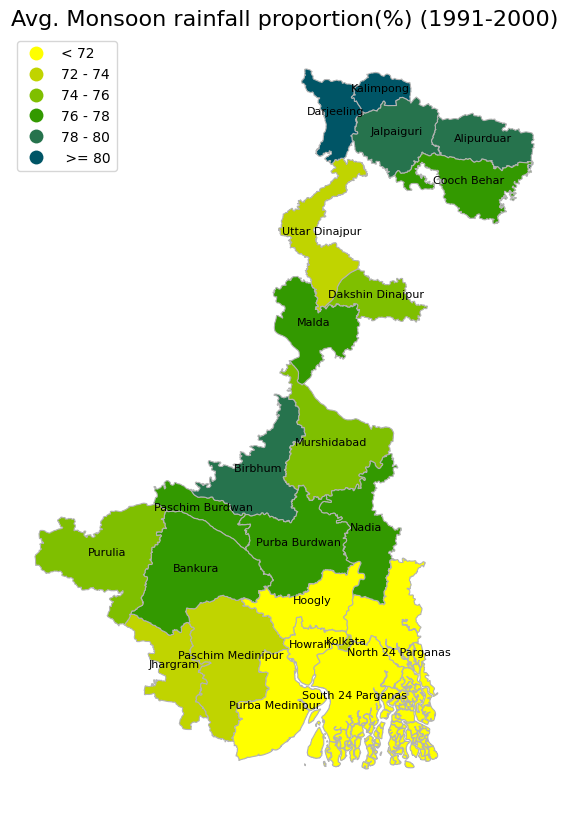} &
        \includegraphics[width=0.48\linewidth]{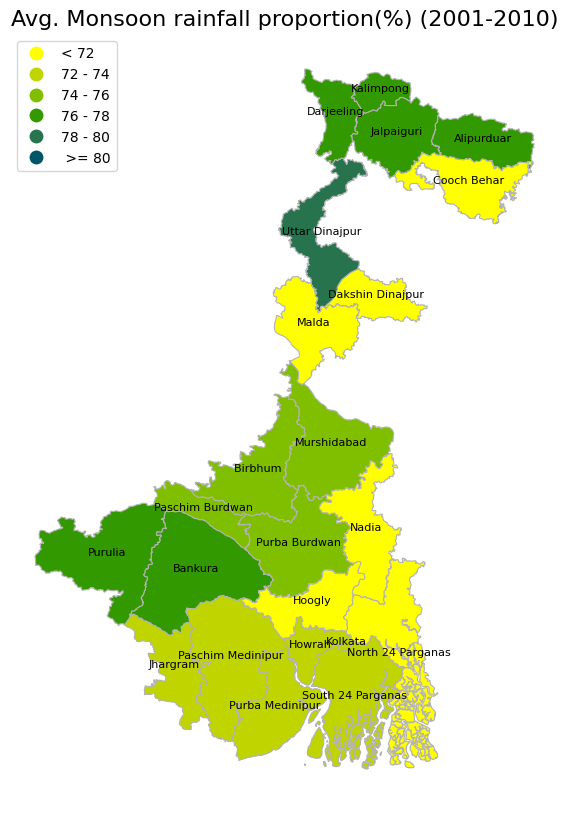} \\
    \end{tabular}
    \caption{Spatial distribution of the proportion of annual rainfall received during the monsoon season across West Bengal over four decades (1971–2010), shown separately for the periods 1971–1980, 1981–1990, 1991–2000, and 2001–2010.}
    \label{fig:map_avg_yearly_monsoon_prop}
\end{figure}

\subsubsection{Decadal Trends in Monsoon Rainfall (1971–2010)}
This section analyzes the characteristics of monsoon rainfall over the past four decades, focusing on two key metrics: the average total monsoon rainfall and the proportion of annual rainfall that occurs during the monsoon season (June to September). The analysis reveals significant spatial and temporal shifts in the state's most critical rainfall period.


\begin{figure}[h!]
    \centering
    \begin{tabular}{cc}
        \includegraphics[width=0.48\linewidth]{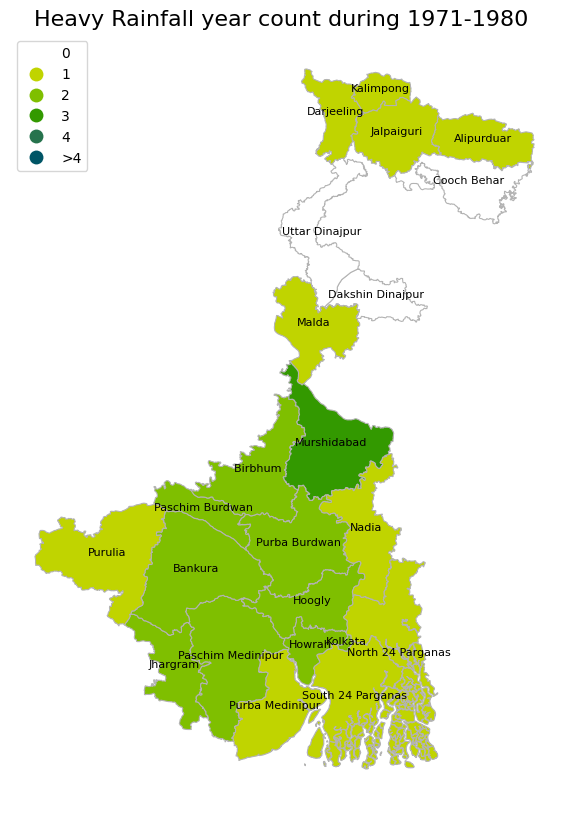} &
        \includegraphics[width=0.48\linewidth]{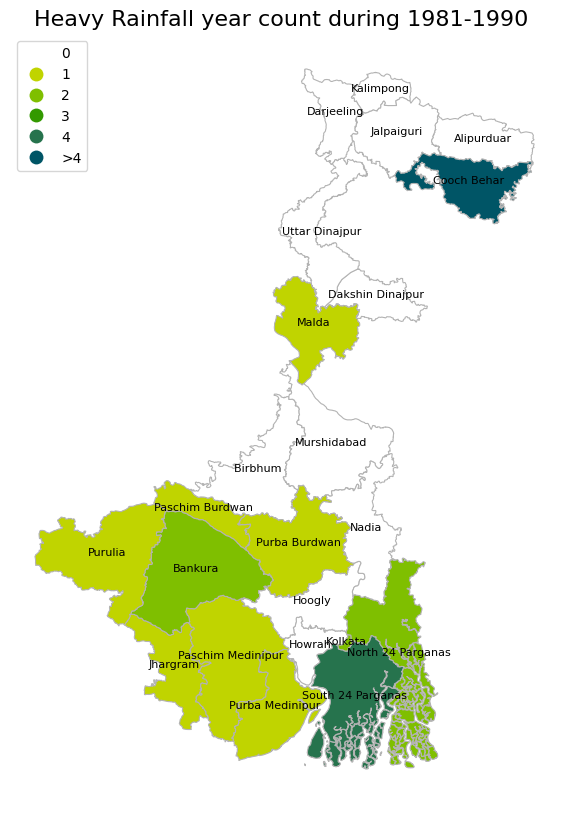} \\
        \includegraphics[width=0.48\linewidth]{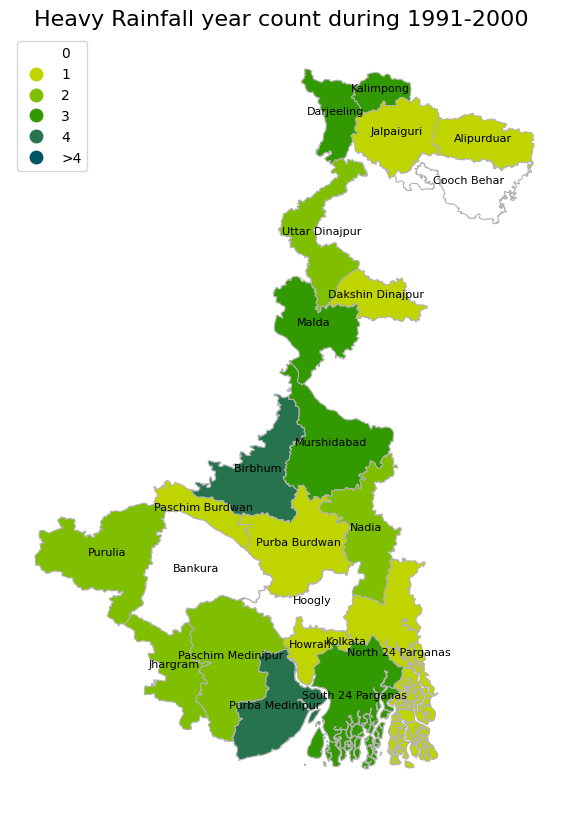} &
        \includegraphics[width=0.48\linewidth]{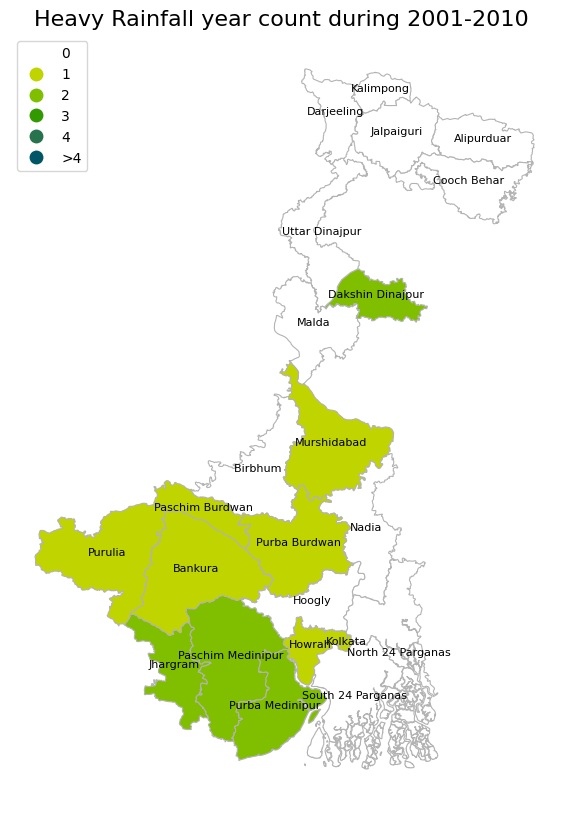} \\
    \end{tabular}
    \caption{Spatial distribution of the number of high-rainfall years per decade across West Bengal over four consecutive periods (1971–2010), shown separately for 1971–1980, 1981–1990, 1991–2000, and 2001–2010.}
    \label{fig:map_heavy_rf}
\end{figure}


\textbf{Average Monsoon Rainfall:}
The maps in Fig.~\ref{fig:map_avg_yearly_monsoon} showing average yearly monsoon rainfall reveal a distinct regional contrast. Districts in the northern hills, such as Darjeeling, Jalpaiguri, and Cooch Behar, consistently receive over 1800 mm of monsoon rainfall across all decades. This highlights their stable orographic advantage and strong dependence on the monsoon. In contrast, districts in the central and eastern plains, like Nadia and Murshidabad, show an "emerging rainfall decline", with average monsoon rainfall gradually decreasing from 1200–1400 mm in the 1970s to below 1200 mm by the 2000s, suggesting a weakening monsoon core in this region. The southwest districts of West Midnapore, Purulia, and Bankura maintain a relatively stable but moderate rainfall, showing no significant positive trend. By the 2001–2010 decade, a distinct polarization emerges, with the north retaining high rainfall while central and eastern districts experience a decline, suggesting a growing intra-state rainfall disparity.

\textbf{Monsoon Proportional Contribution}
The analysis of the proportion (Fig.~\ref{fig:map_avg_yearly_monsoon_prop})
 of yearly rainfall contributed by  monsoon  shows a "rising monsoon dependence" from the 1970s to the 1990s, peaking in most districts with monsoon share exceeding 76\%. North Bengal consistently shows a  high dependence, with its monsoon share exceeding 80\% by the 1990s. In South Bengal, the dependence is more balanced, with many districts retaining a monsoon share below 74\%. The 1990s acted as a "turning point," marking the strongest monsoon concentration and a reduction in the spread of seasonal rainfall. However, in the 2000s, there is a "slight fall in monsoon share",  especially in North Bengal, hinting at increasing rainfall variability and shifting climate patterns. This long-term trend, a transition from rising dependence to a subtle decline in the 2000s, suggests the need for adaptive water management and crop planning to address the changing climate.

\begin{figure}[h!]
    \centering
    \begin{tabular}{cc}
        \includegraphics[width=0.48\linewidth]{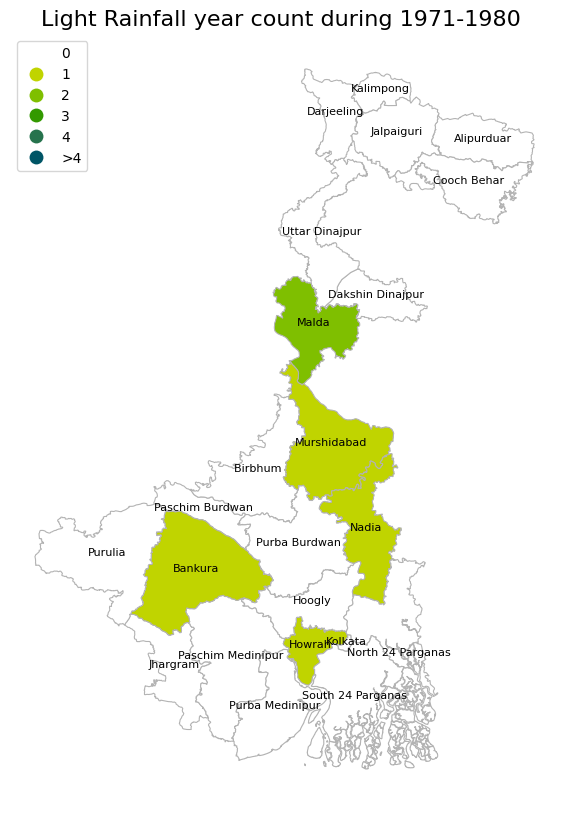} &
        \includegraphics[width=0.48\linewidth]{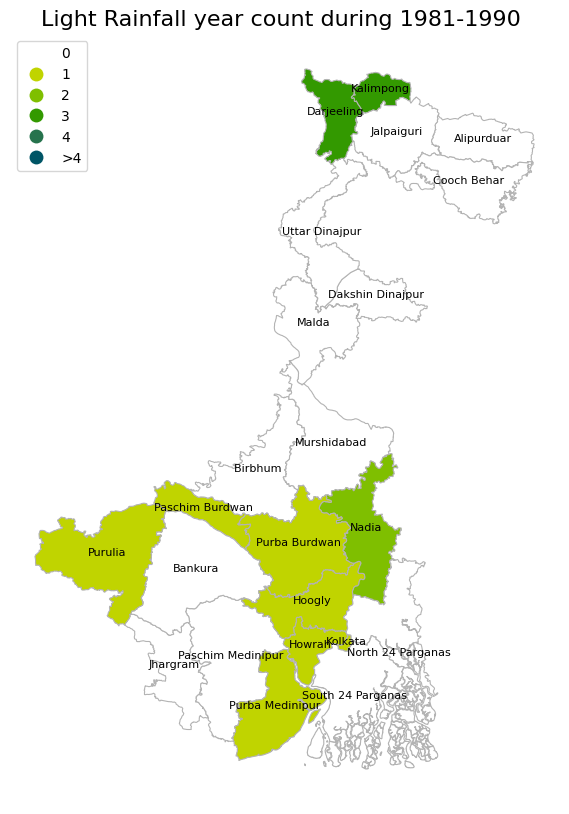} \\
        \includegraphics[width=0.48\linewidth]{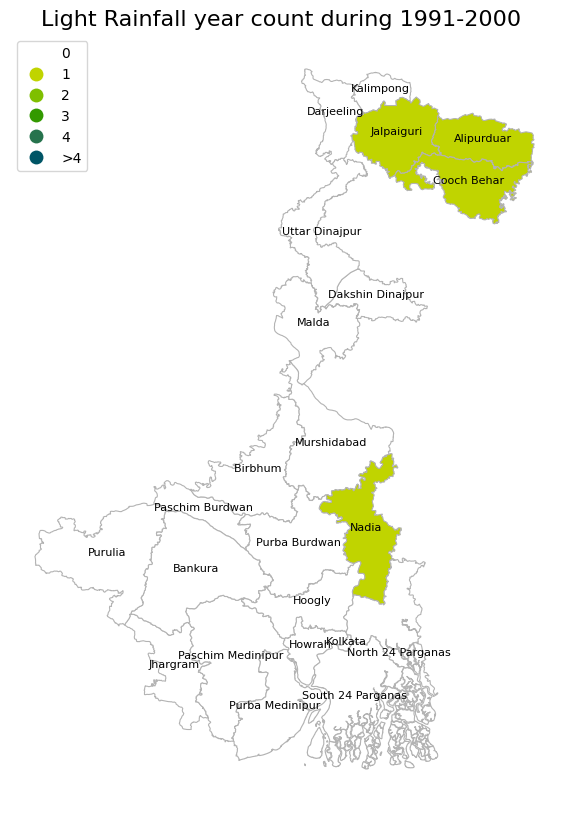} &
        \includegraphics[width=0.48\linewidth]{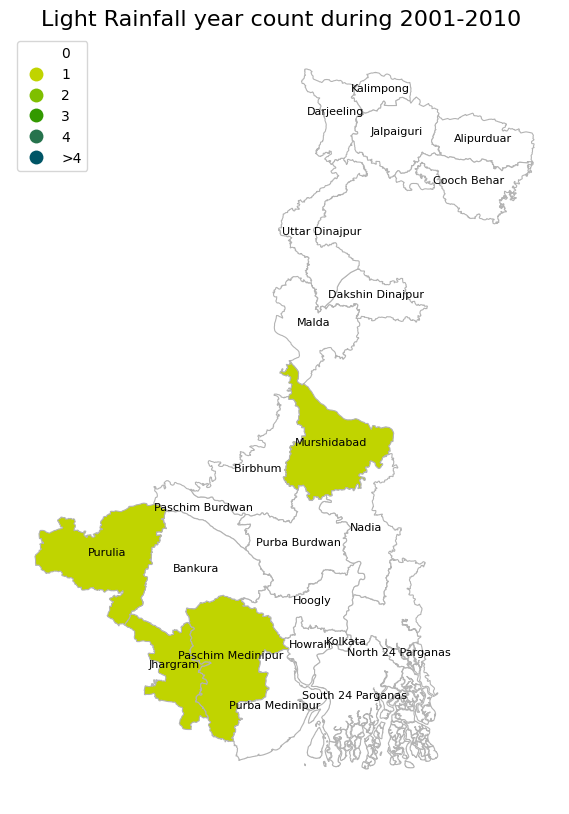} \\
    \end{tabular}
    \caption{Spatial distribution of the number of light-rainfall years per decade across West Bengal over four consecutive periods (1971–2010), shown separately for 1971–1980, 1981–1990, 1991–2000, and 2001–2010.}
    \label{fig:map_light_rf}
\end{figure}

\subsubsection{Analysis of Extreme Rainfall Events}
To gain a more nuanced understanding of rainfall variability beyond simple averages, we analyzed the frequency of extreme rainfall years—both heavy and light—for each district. Our methodology is based on the Standardized Precipitation Index (SPI), a widely used metric for detecting and quantifying meteorological drought and surplus conditions.

\textbf{Methodology for Event Detection}

We first established a baseline using the average and variance of annual rainfall for each district from 1900 to 1970. This period serves as a reference for normalcy, against which subsequent decades are compared. The SPI for any given year is then calculated as the number of standard deviations the annual rainfall of that year deviates from the baseline mean.

\begin{equation}
\text{SPI} = \frac{x - \mu}{\sigma}
\end{equation}

Here, $x$ is the annual rainfall, $\mu$ is the long-term average annual rainfall (1900-1970), and $\sigma$ is the long-term standard deviation of annual rainfall (1900-1970).

Using this approach, we defined:
\begin{itemize}
    \item \textbf{Heavy Rainfall Year}: A year where the SPI is greater than 1.65, which corresponds to the 95\textsuperscript{th} percentile of the rainfall distribution.
    \item \textbf{Light Rainfall Year}: A year where the SPI is less than -1.65, corresponding to the 5\textsuperscript{th} percentile.
\end{itemize}

For each of the four decades (1971–1980, 1981–1990, 1991–2000, and 2001–2010), we counted the number of heavy and light rainfall years for each district. The results are visualized using color-coded maps.



\textbf{Insights and Decadal Trends}

The analysis reveals distinct patterns in the frequency of extreme rainfall events across the decades.

For heavy rainfall events as in Fig.~\ref{fig:map_heavy_rf}, the 1980s and 1990s saw a significant "spike in heavy years," particularly in North Bengal and coastal districts. The 1990s were marked by the most "widespread intense rainfall" across the state, indicating a period of high hydro-climatic variability. However, this trend saw a "significant decline" in the 2000s, with most districts experiencing only a few heavy rainfall events. This post-2000 dip may be a sign of changing monsoon patterns or a shift in rainfall concentration.

Regarding light rainfall or drought years (Fig.~\ref{fig:map_light_rf}), the 1970s showed "scattered droughts" with moderate frequency in central districts. The 1980s saw an "increase in drought-prone years" in the northern hills (e.g., Darjeeling) and southwestern districts (e.g., Purulia). The 1990s, despite the heavy rainfall spike, also showed a "sharp decline in light years," suggesting fewer drought-like conditions and a more balanced rainfall distribution. However, the 2000s saw a rise in light rainfall years again, with a "sparse drought signals overall," but Bankura and Purulia continue to show persistent vulnerability to these conditions across all decades. The observed patterns suggest that while West Bengal experienced a period of high variability with both heavy rainfall and drought years in the 1980s and 1990s, the subsequent decade appears to have been relatively more stable, albeit with a trend towards decreasing heavy rainfall events.

\subsubsection{Spatial Dependence of Rainfall Characteristics}

To investigate how the similarity of rainfall patterns changes with geographical distance, we conducted a spatial analysis based on pairwise comparisons of districts. For this analysis, each of the 19 districts in West Bengal was assigned a representative latitude and longitude. This was determined by taking the average latitude and longitude of all meteorological observatories within that district. As the districts are largely convex in shape, this serves as a robust central point.

\textbf{Methodology}

For every unique pair of districts, the geographical distance between their representative points was calculated in kilometers using the `geopy.distance.distance` function. This provided a total of ${19 \choose 2}$ unique distance pairs. For each of these pairs, we then calculated the Pearson correlation coefficient between the time series of various rainfall-related metrics. This approach allowed us to visualize the relationship between the pairwise correlation of a given metric and the pairwise distance between districts. A LOESS regression curve was fitted to each plot to illustrate the general trend.

\textbf{Key Findings}

\begin{figure*}[htbp]
    \centering
    \includegraphics[width=\textwidth]{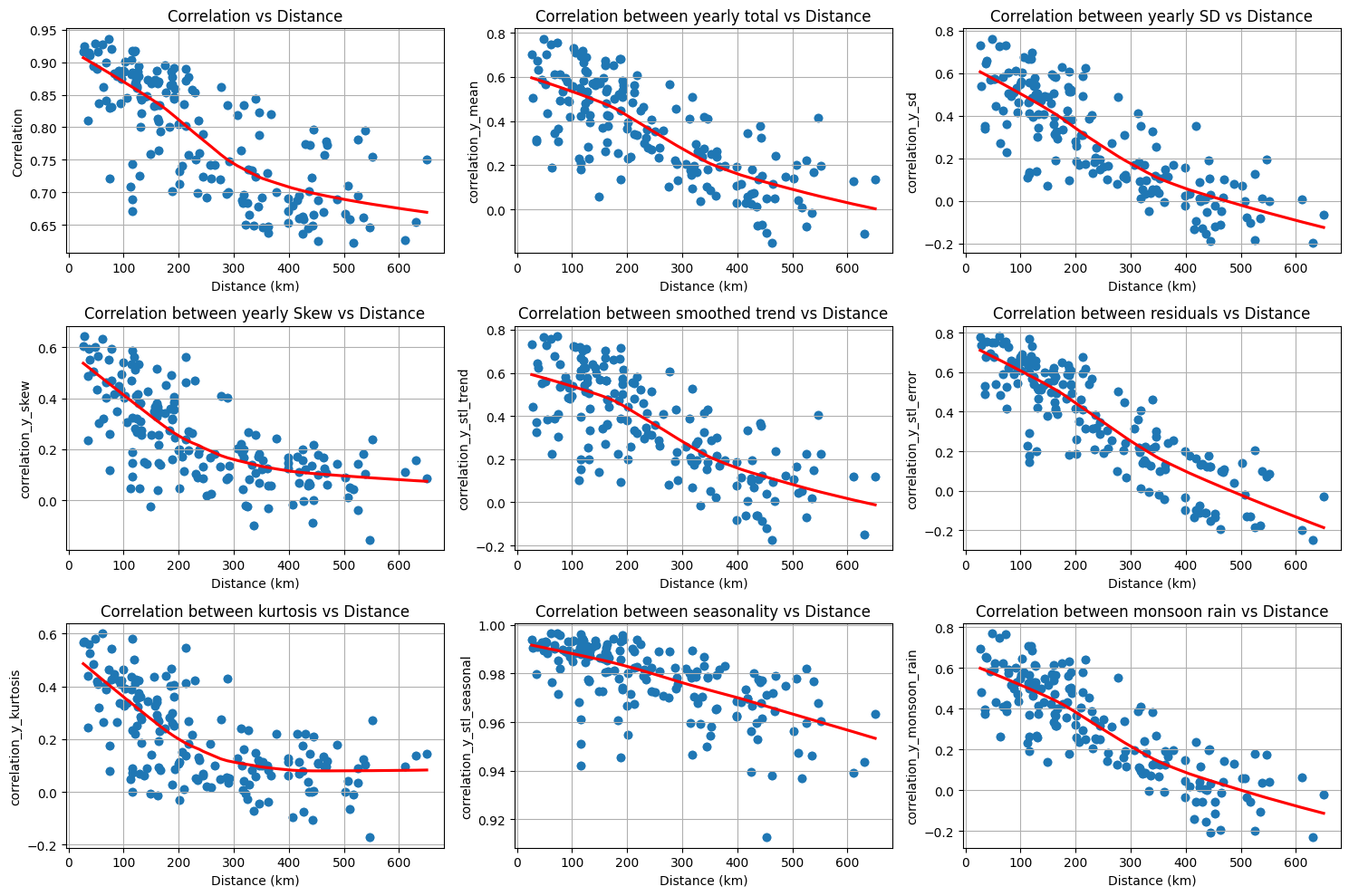}
    \caption{
Spatial decay of rainfall similarity across West Bengal. 
Each scatter plot shows the Pearson correlation of a specific rainfall characteristic 
(e.g., monthly totals, yearly aggregates, higher-order moments, STL components) 
between district pairs, plotted against their geographical separation. 
A clear inverse relationship is observed across all metrics, confirming that 
rainfall dynamics exhibit strong spatial dependence that weakens with distance. 
This justifies the use of distance-based neighbourhood structures in the proposed models.}

    \label{fig:wide_figure}
\end{figure*}

The analysis, as visualized in the plots (Figure 1), consistently demonstrates a "strong spatial decay" in the correlation of rainfall characteristics as the distance between districts increases.

\begin{enumerate}
    \item \textbf{Monthly Rainfall}: The correlation of monthly rainfall shows a sharp inverse relationship with distance. Nearby districts (within 100-150 km) exhibit a high correlation ($>0.9$), which then drops significantly beyond 300 km. This indicates a strong local coupling of monthly rainfall dynamics.
    \item \textbf{Yearly Mean and Standard Deviation}: This decaying spatial dependence is also observed in annual rainfall aggregates. The plots show that the pairwise correlations of both yearly mean rainfall and yearly standard deviation also decrease as the distance between districts increases. This supports the idea that the intensity and variability of rainfall are regionally clustered.
    \item \textbf{Higher-Order Moments}: The correlations for higher-order moments, namely yearly skewness and yearly kurtosis, show weaker spatial structure and a more rapid decay. This suggests that rainfall asymmetry and extreme events are more localized and influenced by microclimatic or topographic factors.
    \item \textbf{STL Decomposition Components}: We further decomposed the monthly time series for each district using the Seasonal-Trend decomposition procedure based on Loess (STL). The resulting components provide insights into the underlying drivers of rainfall patterns.
    \begin{itemize}
        \item \textbf{Smoothed Trend}: The correlation of the smoothed trend component decays gradually with distance. This reinforces the idea that long-term climatic drivers and multi-year monsoon cycles affect neighboring districts in a similar, geographically dependent manner.
        \item \textbf{Seasonality}: The seasonal component exhibits a very strong spatial correlation, which decays only slightly with distance. This is expected, as the annual monsoon cycle is a dominant, large-scale meteorological phenomenon that affects the entire region in a synchronized manner.
        \item \textbf{Residuals}: The residual component, which represents short-term or irregular fluctuations, shows the lowest correlation with distance. This implies that such random fluctuations are not spatially coherent.
    \end{itemize}
    \item \textbf{Monsoon Rainfall}: The yearly monsoon rainfall, which is a key aggregate, shows a strong regional coherence. Its correlation remains high for distances up to 200 km but drops off significantly beyond that. This suggests that the monsoon activity is spatially coherent within regions, but its influence has a definite limit beyond which local effects dominate.
\end{enumerate}

Overall, these consistent negative spatial correlation gradients across multiple rainfall metrics strongly validate a modeling strategy that accounts for regional dependencies, such as k-nearest neighbor or other distance-based neighborhood models, for spatio-temporal rainfall forecasting.

\section{Model Methodology}

In this section, we describe the forecasting methodologies evaluated in the study. Three models are considered, ranging from simple baselines to the proposed framework. The first is a seasonal naïve model, serving as a benchmark for basic seasonal patterns. The second, a Spatio-Temporal Lag Model (STLM), employs multi-layer perceptrons with both own and neighboring district lags. The final model, the Hierarchical Spatio-Temporal Model (HSTM), introduces yearly aggregated features with smoothing, integrated into an MLP-based monthly forecasting framework.
\subsection{Baseline Model: Seasonal Naïve}

\subsubsection{Model specification}
The seasonal naïve model forecasts each future observation by repeating the value observed in the same month of the previous year. Let $m=12$ denote the seasonal period (months in a year). For district $d$, with monthly rainfall series $\{y_{d,t}\}_{t=1}^T$, and a forecasting origin at time $T_0$, the forecast for horizon $h$ is
\begin{equation}
\hat{y}_{d,T_0+h} = y_{d,T_0+h-m}.
\end{equation}
This rule directly exploits the strong annual seasonality in rainfall and requires no parameters or estimation.

\subsubsection{Procedure}
The forecasting process is applied independently to each of the 19 district time series:
\begin{enumerate}
    \item Split each series into a training set $\{y_{d,1},\dots,y_{d,T_0}\}$ and a test set $\{y_{d,T_0+1},\dots,y_{d,T}\}$.
    \item Generate forecasts $\hat{y}_{d,T_0+1},\dots,\hat{y}_{d,T}$ using the seasonal naïve rule.
    \item Compute forecast errors $e_{d,h} = y_{d,T_0+h} - \hat{y}_{d,T_0+h}$.
    \item Store forecasts and errors for further evaluation and visualization.
\end{enumerate}

\subsubsection{Remark 1}

In this study, the training set for each district consists of monthly rainfall data from January 1900 to December 2010, while the test set spans January 2011 to December 2019. The performance results of the model are presented in Section VII.

\subsubsection{Remark 2}
The seasonal naïve model assumes a stable, repeating seasonal cycle, making it appropriate as a simple benchmark for monthly rainfall forecasting. It does not account for long-term trends, interannual variability, or nonlinear dependencies. Its main role in this study is to provide a transparent baseline against which more complex spatio-temporal models can be compared.

\subsection{Benchmark Model: Spatio-Temporal Lag Model (STLM)}

\subsubsection{Model Intuition and Structure}

The Spatio-Temporal Lag Model (STLM) extends a simple autoregressive idea by allowing each district’s rainfall to depend not only on its own past values but also on the recent history of its neighboring districts. The aim is to capture both \textit{temporal persistence} (rainfall patterns repeating across months within a district) and \textit{spatial dependence} (rainfall often co-moving across nearby districts).

Formally, let $y_{d,t}$ denote the rainfall in district $d$ at month $t$. The forecast for month $t$ is constructed from two components:

\begin{enumerate}
    \item \textbf{Own lags:} the past $p$ values of district $d$,
    \[
    y_{d,t-1},\; y_{d,t-2},\; \dots,\; y_{d,t-p}.
    \]

    \item \textbf{Neighbor lags:} for each of the $k$ nearest neighbors of district $d$, we include the past $q$ values. If the neighbors are indexed as $n_1,\dots,n_k$, then the neighbor inputs are
    \[
    y_{n_j,t-1},\; y_{n_j,t-2},\; \dots,\; y_{n_j,t-q}, \qquad j=1,\dots,k.
    \]
\end{enumerate}

Together, the input vector at time $t$ for district $d$ is

\[
\mathbf{X}_{d,t} =
[
y_{d,t-1},\dots,y_{d,t-p};  
y_{n_1,t-1},\dots,y_{n_1,t-q};\; \ldots;\; \]
\[ y_{n_k,t-1},\dots,y_{n_k,t-q}
]^\top. \]

This input is fed into a multilayer perceptron (MLP), denoted $f_\theta(\cdot)$, which learns nonlinear relationships between inputs and rainfall. The one-step-ahead forecast is then
\begin{equation}
\hat{y}_{d,t} = f_\theta(\mathbf{X}_{d,t}),
\end{equation}
where $\theta$ are the trainable network parameters (weights and biases).

\textit{Intuitively:}
\begin{itemize}
    \item $p$ controls how much temporal memory of the district is used.
    \item $k$ controls how much spatial information is borrowed from neighbors.
    \item $q$ controls how deep in time we look for each neighbor.
\end{itemize}

By combining these with a flexible MLP, the STLM can approximate complex spatio-temporal dependencies beyond what linear models capture.

\subsubsection{Parameters and Search Space}

The STLM involves two broad categories of parameters: \textit{structural parameters}, which determine the input representation, and \textit{training parameters}, which govern how the neural network is fitted. Both play crucial roles in balancing model flexibility and predictive accuracy.

\paragraph{ Structural parameters (input-related)}
These control how the spatio-temporal input vector $\mathbf{X}_{d,t}$ is formed.
\begin{itemize}
    \item \textbf{Own lags ($p$):} the number of past monthly values of the same district included. Larger $p$ allows the model to capture longer temporal memory, but reduces the effective training size.
    \item \textbf{Number of neighbors ($k$):} how many neighboring districts are considered. A larger $k$ increases spatial information but also input dimensionality, which can make the model harder to train.
    \item \textbf{Neighbor lags ($q$):} the number of past months considered for each neighbor. Larger $q$ captures more detailed neighbor history but again enlarges the feature space.
\end{itemize}
Together, these determine the input dimension:
\begin{equation}
\dim(\mathbf{X}_{d,t}) = p + k \cdot q.
\end{equation}

\paragraph{ Training parameters (MLP and optimization)}
These affect how the MLP $f_\theta(\cdot)$ fits the given inputs.
\begin{itemize}
    \item \textbf{Learning rate ($\eta$):} This controls the step size of gradient updates; too small slows convergence, too large may cause instability.
    \item \textbf{Regularization parameter ($\alpha$):} The inclusion of neighbour information substantially increases the input dimensionality, as both the number of neighbours ($k$) and the number of lags ($q$) grow. Moreover, when additional districts are incorporated, including all their lagged values can further expand the feature space and raise the risk of overfitting. To address this, feature selection and regularization are essential. The L1 penalty (LASSO) is particularly well-suited in this context, as it simultaneously shrinks coefficients and eliminates irrelevant predictors, thereby controlling model complexity while improving generalization.

    \item \textbf{Architecture:} The number of hidden layers and number of units per layer. More layers/units increase model capacity but risk overfitting.
    \item \textbf{Epochs and batch size:} Its define how many passes the optimizer makes through the data and the sample size per gradient update. Batch size is fixed (e.g., 64), while epochs are tuned with early stopping.
\end{itemize}

\begin{table*}[htbp]
\centering
\caption{Final model specifications of STLM after cross-validation.}
\label{tab:stlm_specs}
\begin{tabular}{lcccccccc}
\hline
District & $p$ & $k$ & $q$ & Layers & Units & $\eta$ & $\alpha$ & Epochs \\
\hline
BANKURA         & 120 & 4 & 2 & 2 & (8,4)  & $10^{-4}$  & $10^{-5}$ & 60 \\
BIRBHUM         & 140 & 5 & 1 & 2 & (10,5) & $10^{-3}$  & $10^{-2}$ & 60 \\
BURDWAN         & 180 & 2 & 3 & 2 & (2,2)  & $10^{-2}$  & $10^{-5}$ & 30 \\
COOCH BEHAR     & 160 & 4 & 1 & 2 & (8,4)  & $10^{-4}$  & $10^{-3}$ & 80 \\
DARJEELING      & 100 & 3 & 4 & 2 & (3,2)  & $10^{-3}$  & $10^{-2}$ & 30 \\
HOOGLY          & 120 & 4 & 4 & 2 & (7,6)  & $10^{-1}$  & $10^{-4}$ & 40 \\
HOWRAH          & 120 & 4 & 2 & 2 & (4,4)  & $10^{-2}$  & $10^{-2}$ & 50 \\
JALPAIGURI      & 140 & 2 & 1 & 2 & (6,4)  & $10^{-2}$  & $10^{-4}$ & 40 \\
MALDA           & 160 & 5 & 2 & 2 & (2,2)  & $10^{-2}$  & $10^{-2}$ & 30 \\
MANBHUM PURULIA & 100 & 2 & 5 & 2 & (3,2)  & $10^{-2}$  & $10^{-2}$ & 50 \\
EAST MIDNAPORE  & 140 & 3 & 1 & 2 & (5,5)  & $10^{-4}$  & $10^{-1}$ & 40 \\
MURSHIDABAD     & 140 & 4 & 2 & 2 & (4,4)  & $10^{-4}$  & $10^{-3}$ & 50 \\
NADIA           & 120 & 2 & 4 & 2 & (7,6)  & $10^{-3}$  & $10^{-3}$ & 70 \\
24 PARGANAS N   & 160 & 2 & 3 & 2 & (4,2)  & $10^{-1}$  & $10^{-3}$ & 30 \\
24 PARGANAS S   & 180 & 4 & 4 & 2 & (3,2)  & $10^{-4}$  & $10^{-2}$ & 40 \\
DINAJPUR NORTH  & 160 & 5 & 2 & 2 & (4,4)  & $10^{-2}$  & $10^{-1}$ & 30 \\
DINAJPUR SOUTH  & 120 & 3 & 4 & 3 & (3,3)  & $10^{-4}$  & $10^{-5}$ & 30 \\
WEST MIDNAPORE  & 140 & 1 & 1 & 2 & (2,2)  & $10^{-5}$  & $10^{-3}$ & 60 \\
KOLKATA         & 120 & 2 & 1 & 2 & (2,2)  & $10^{-4}$  & $10^{-4}$ & 40 \\
\hline
\end{tabular}
\end{table*}

\paragraph{Search space for optimization}

The optimization procedure involves a random search approach that considers 2000 random samples without replacement from the search space. To identify suitable parameter values, we define a grid of candidate configurations for cross-validation. Given the relatively small dataset (approximately 1,000 observations) and the tendency of MLPs to become increasingly complex with deeper architectures, the search is restricted to simpler networks with at most two hidden layers. To further reduce the search space and streamline the tuning procedure, the batch size is fixed at 32. The resulting ranges (which may be refined in future work) are as follows.
\begin{align*}
p &\in \{80, 100, 120, 140, 160, 180 \}, \\
k &\in \{0,1, 2,3,4, 5\}, \\
q &\in \{5, 10, 20, 40 \}, \\
\text{Hidden layers} &= 2 \ (\text{fixed}), \\
\text{Units per layer} &\in \{2, 3, 4,5, 6, 7, 8, 9, 10\}, \\
\eta &\in \{10^{-5},10^{-4},10^{-3},10^{-2}, 10^{-1}\}, \\
\alpha &\in \{10^{-5},10^{-4},10^{-3},10^{-2}, 10^{-1}\}, \\
\text{Batch size} &= 32 \ (\text{fixed}), \\
\text{Epochs} &\in \{20, 30, ..., 100\} \ (\text{with early stopping}).
\end{align*}

The search space for $p$ (own lags) is chosen to be relatively large, while that for $k$ 
(number of neighbours) is restricted to $\{1,2,3,4,5\}$. Preliminary experiments indicated 
that in long-term forecasting (over 100 months), small values of $p$ often lead to rapid 
error accumulation and, in some cases, unstable forecasts. In contrast, increasing the number 
of neighbours beyond a small set had only marginal benefit compared to using fewer neighbours 
with deeper lag information. Furthermore, given the limited dataset, simpler networks 
consistently outperformed larger ones, leading us to fix the number of hidden layers at two.

\subsubsection{Time-Series Cross-Validation and Parameter Tuning}

To select the optimal parameters of the STLM, we use \textit{time-series cross-validation} (TCV) with an expanding window design. Unlike standard cross-validation for cross-sectional data, time-series CV respects the temporal order: training is always on earlier observations and validation is on subsequent periods, avoiding any look-ahead bias.

\textit{a) Expanding window setup:} 
Let the training set for each district cover January 1900 to December 2010, with $T_0$ total months. We split this into $K=5$ folds. For fold $i$, we define:
\begin{itemize}
    \item Training set: from the start of the series up to month $s_i - 1$,
    \item Validation set: the next $h_{\text{val}}$ months, starting from $s_i$,
\end{itemize}
where
\begin{equation}
s_i = T_0 - K \cdot h_{\text{val}} + (i-1)h_{\text{val}} + 1, 
\quad i=1,\dots,K,
\end{equation}
and $h_{\text{val}}=120$ months (10 years). This design ensures that each fold uses an expanding training window with a rolling validation block, allowing forecasts to be evaluated over longer horizons. Such evaluation is important, as model architectures that perform well for short-term prediction may not generalize effectively to long-term forecasting.

\textit{c) Selection rule and metrics}

For each fold, the data are divided into a training period and a corresponding validation period. During the training phase, hyperparameters are selected independently for each district. Specifically, for district $d$, a random configuration
\[
\theta_d = (p,\ q,\ k,\ \text{number of hidden layers},\ \eta,\ \alpha,\ \text{epochs})
\]
is sampled from the predefined search space. A complete hyperparameter draw for the full multi-district system is therefore given by
\[
\theta = (\theta_1,\ \theta_2,\ \dots,\ \theta_D),
\]
where $D$ denotes the total number of districts (here, $D = 19$). For each sampled $\theta$, the model is jointly fitted using the district-specific parameter choices. A total of 2000 such samples are evaluated as part of the randomized search procedure.

Forecasts for the validation set are then generated in a recursive multi-output manner: since the prediction for any district depends not only on its own lagged values but also on the lagged values of its neighbouring districts, all districts are forecasted jointly. Specifically, a one-step-ahead prediction is first produced for every district at time $t$, and these forecasts are then fed back as inputs to generate predictions for time $t+1$. This procedure is repeated iteratively until the entire validation horizon is covered.

For each district $d$ in the validation period, we obtain both the actual monthly rainfall series $\{y_{d,t}\}$ and the corresponding forecasted series $\{\hat{y}_{d,t}\}$. The district-wise prediction error is measured using the Root Mean Squared Error (RMSE):
\begin{equation}
\mathrm{RMSE}_d = \sqrt{\frac{1}{K} \sum_{t=1}^{K} (y_{d,t} - \hat{y}_{d,t})^2},
\end{equation}
where $K$ denotes the number of months in the validation period.

However, optimizing performance separately for each district is not appropriate in this setting because the forecasts are interdependent: errors in one district propagate to others through the lag structure. Therefore, instead of minimizing $\mathrm{RMSE}_d$ individually, we assess forecasting performance collectively across all districts. To account for the difference in rainfall scale across districts, we use the Normalized Root Mean Squared Error (NRMSE), defined as
\begin{equation}
\mathrm{NRMSE}_d = \frac{\mathrm{RMSE}_d}{\mathrm{SD}_d},
\end{equation}
where $\mathrm{SD}_d$ is the standard deviation of the observed rainfall series for district $d$ over the validation period.

For each fold $i$, the average NRMSE across districts is computed as
\begin{equation}
\overline{\mathrm{NRMSE}}^{(i)} = \frac{1}{D} \sum_{d=1}^{D} \mathrm{NRMSE}_d^{(i)}, \qquad D = 19.
\end{equation}
Finally, the overall performance metric used for model selection is obtained by averaging across folds:
\begin{equation}
\overline{\mathrm{NRMSE}} = \frac{1}{K} \sum_{i=1}^{K} \overline{\mathrm{NRMSE}}^{(i)}.
\end{equation}

This metric captures the joint predictive accuracy of all district-level series while accounting for both scale differences and interdependence, making it a robust criterion for model selection in the spatio-temporal setting.

\subsubsection{Final Model Specifications}

After completing the time-series cross-validation procedure, the optimal parameter
$\theta^\star =  (\theta_1^\star,\ \theta_2^\star,\ \dots,\ \theta_D^\star),
$ is selected that contains the parameter configuration for each district. These specifications include both 
the structural parameters that determine the input design and the model parameters 
that control training. Documenting the final configuration ensures transparency 
and reproducibility.

The final optimized parameters for all districts are presented in Table~\ref{tab:stlm_specs}. 
The first block shows the search ranges used during cross-validation, while the 
second block reports the selected values for each district.

\textit{Remarks:}
The table provides a compact summary of the optimized configurations. 
Although the search spaces were common across districts, the selected values 
may differ due to heterogeneity in rainfall dynamics. All models were trained 
with a fixed batch size of 32, ReLU activations, and the Adam optimizer.

\subsection{Hierarchical Spatio-Temporal Model (HSTM)}

\subsubsection{Model intuition and overview}

The Hierarchical Spatio-Temporal Model (HSTM) extends the STLM by introducing a two-stage pipeline that explicitly models slowly varying annual characteristics and uses those forecasts as auxiliary inputs into a monthly forecasting network. The intuition is straightforward: some aspects of rainfall dynamics (annual total, distribution across quarters, variability and extremes) evolve at a much slower time scale than month-to-month fluctuations. By forecasting these yearly summaries using a lightweight spatio-temporal regression and then conditioning a monthly MLP on the predicted yearly behavior, the HSTM combines stability from aggregated signals with the fine temporal resolution captured by the monthly MLP. This reduces long-horizon instability, error accumulation and supplies the monthly model with high-level climatic context unavailable in short lag windows. This approach is particularly useful in smaller data sets where training deep neural networks are not feasible. 

\subsubsection{Notation and Yearly Feature Construction}

Let $y_{d,t}$ denote the monthly rainfall in district $d \in \{1,\dots,19\}$ at month $t$. 
For each year $T$ and district $d$, we collect the 12 consecutive monthly values
\[
\{m_{d,T,1}, m_{d,T,2}, \dots, m_{d,T,12}\},
\]
where $m_{d,T,j}$ denotes rainfall in month $j$ ($j=1$ = January, \dots, $j=12$ = December) of year $T$.  
From these 12 values we construct a set of yearly features, each capturing a distinct characteristic of the annual rainfall profile.

\textit{a) Yearly Total:}
\begin{equation}
\text{Total}_{d,T} = \sum_{j=1}^{12} m_{d,T,j}
\end{equation}
The yearly total represents the overall annual water availability. It is the most direct indicator of how wet or dry a year is, serving as the baseline against which all other features are interpreted. Unlike relative measures such as entropy or quarterly proportions, the yearly total carries absolute magnitude information, which is essential for understanding interannual rainfall variability and long-term water balance.

\textit{b) Monsoon Total:}
\begin{equation}
\text{MonsoonTotal}_{d,T} = \sum_{j=6}^{9} m_{d,T,j}
\end{equation}
The monsoon total isolates rainfall during the June--September southwest monsoon season, which is the dominant driver of agriculture and hydrology in West Bengal. While the yearly total gives the overall water budget, the monsoon total captures the intensity of the season most critical for cropping cycles and reservoir inflows. This feature allows us to distinguish years with similar annual totals but different seasonal distributions.

\textit{c) Entropy:}
\begin{equation}
E_{d,T} = -\frac{1}{\log 12}\sum_{j=1}^{12} p_{d,T,j}\,\log p_{d,T,j}, 
\quad p_{d,T,j}=\frac{m_{d,T,j}}{\text{Total}_{d,T}}
\end{equation}
Entropy measures the evenness of monthly rainfall distribution within a year. A high entropy value indicates that rainfall is spread relatively evenly across months, whereas a low value signifies concentration in a few months. This feature complements the magnitude-based indicators by characterizing the shape of the rainfall distribution, thereby distinguishing between years that may have the same total but very different seasonal structures.

\textit{d) Standard Deviation:}
\begin{equation}
\text{SD}_{d,T} = \sqrt{\frac{1}{12}\sum_{j=1}^{12}\big(m_{d,T,j}-\bar{m}_{d,T}\big)^2}, 
\\
\quad \bar{m}_{d,T}=\frac{1}{12}\sum_{j=1}^{12} m_{d,T,j}
\end{equation}
The standard deviation quantifies the variability in monthly rainfall magnitudes. Unlike entropy, which is scale-free, SD preserves the absolute dimension of variability, highlighting whether rainfall is highly concentrated in a few intense months or more balanced across the year. It is particularly useful for identifying years dominated by sharp monsoon peaks, which may not be reflected in entropy alone.

\textit{e) Yearly Centroid:}
\begin{equation}
C_{d,T} = \sum_{j=1}^{12} j \cdot p_{d,T,j}
\end{equation}

The yearly centroid provides a continuous measure of the timing of rainfall within the year, analogous to the ``center of mass'' of the monthly distribution. Lower values indicate that rainfall is concentrated earlier (e.g., June--July), while higher values indicate later-season dominance (e.g., August--September). This feature uniquely captures timing shifts in the rainfall regime that magnitude or variability measures cannot reveal.

\textit{f) Maximum Monthly Rainfall:}
\begin{equation}
\text{Max}_{d,T} = \max_{j=1,\dots,12} m_{d,T,j}
\end{equation}
The maximum monthly rainfall captures the most intense rainfall month in a year, serving as a proxy for extremes and potential flood risk. It highlights interannual fluctuations in rainfall peaks, complementing the yearly total by focusing on single-month extremes that can have disproportionate socio-economic and hydrological impacts.

\textit{g) Quarterly Proportions:}
\begin{equation}
Q_{d,T,q} = \frac{\sum_{j=3(q-1)+1}^{3q} m_{d,T,j}}{\text{Total}_{d,T}}, 
\quad q=1,2,3.
\end{equation}
Thus $Q_{1}$ = Jan--Mar, $Q_{2}$ = Apr--Jun, $Q_{3}$ = Jul--Sep, with $Q_4$ omitted since it is redundant ($Q_4 = 1-(Q_1+Q_2+Q_3)$). The quarterly proportions allocate rainfall into coarse seasonal blocks. Unlike entropy or centroid, which provide smooth distributional measures, quarterly proportions allow explicit quantification of how much rainfall is concentrated in distinct seasonal periods, making them useful for agricultural and water management applications.

Beyond the features defined above, several other descriptors could in principle be constructed to summarize yearly rainfall. For instance, the interquartile range (IQR) could serve as a robust variability measure less sensitive to extreme months than the standard deviation, while the coefficient of variation (CV) provides a scale-free index of dispersion. Measures such as the Gini index or Theil index could be used to capture inequality in monthly rainfall contributions, analogous to income inequality metrics. Similarly, higher-order moments like kurtosis could provide information on the sharpness of seasonal peaks, and circular statistics could be applied to refine timing measures. While these alternatives carry interpretive value, we restrict attention to the selected set of features to keep the framework concise, interpretable, and well-aligned with hydrological and agricultural relevance. Including too many overlapping or complex indicators risks redundancy and over-complication without necessarily improving forecasting performance.

\subsubsection{Smoothing and Span Hyperparameter}

The yearly feature time series derived in the previous subsection contain only 111 observations (1900–2010). At this temporal resolution, the series are affected by considerable interannual variability, which may reflect local fluctuations, measurement errors, or small-scale meteorological anomalies. Directly forecasting these unsmoothed features risks overfitting to year-specific noise rather than capturing the underlying long-term dynamics. To mitigate this, we apply \textit{exponential moving average (EMA) smoothing} to each yearly feature series prior to forecasting. The purpose of forecasting the yearly features is not to achieve highly accurate year-level predictions, but rather to extract the broader temporal signal carried in these smoothed series. These forecasts are then incorporated as inputs in the second stage of the model, where an MLP-based monthly forecasting framework combines the yearly features with own lags and neighboring information to improve district-level rainfall predictions.

Formally, let $x_{d,T}$ denote a yearly feature for district $d$ in year $T$. The smoothed 
feature series $F_{d,T}$ is defined recursively as
\begin{equation}
F_{d,T} = \alpha x_{d,T} + (1-\alpha)F_{d,T-1}, \quad 0<\alpha\leq 1,
\end{equation}
with initialization $F_{d,0} = x_{d,0}$. The parameter $\alpha$ is related to the 
\textit{span} $s$ by
\[
\alpha = \frac{2}{s+1}.
\]
Here, $s$ controls the effective degree of smoothing: smaller values of $s$ lead to a larger 
$\alpha$ and hence less smoothing (the series reacts strongly to new observations), while 
larger values of $s$ correspond to a smaller $\alpha$ and hence stronger smoothing (the series 
changes slowly over time).

Each of the yearly feature sets (Total, Monsoon Total, Entropy, Standard Deviation, 
Centroid, Maximum, $Q_1$, $Q_2$, $Q_3$) is smoothed with its own span value. This design 
recognizes that different features exhibit distinct levels of natural variability. For 
example, Yearly Total tends to be highly variable and may benefit from stronger smoothing, 
while Entropy or Centroid evolve more gradually and can be preserved with weaker smoothing. For each yearly feature type (e.g., yearly total, yearly entropy, etc.), there are $D = 19$  corresponding yearly time series, one for each district. Within a given feature set, a single span value is used uniformly across all districts. For example, when smoothing the yearly total series, the same span parameter is applied to all 19 districts; similarly, a common span value is used for smoothing all entropy series.

\textit{Under-smoothing versus Over-smoothing:}  
\begin{itemize}
    \item \textbf{Under-smoothing (small $s$):} Too little smoothing leaves substantial noise 
    in the feature series. Forecasts of these noisy features may deviate strongly from their 
    true underlying dynamics, and when used as auxiliary inputs in the Stage-2 MLP, they can 
    distort the network’s understanding of train--test relationships. In effect, the neural 
    network may interpret random year-to-year fluctuations as signal.  
    \item \textbf{Over-smoothing (large $s$):} Excessive smoothing removes meaningful 
    variability and flattens genuine shifts in the feature time series. Forecasts then become 
    overly smooth, limiting the ability of the downstream model to capture interannual extremes 
    or decadal shifts. Over-smoothed features risk becoming redundant, carrying little 
    additional information beyond long-term averages.
\end{itemize}

\textit{Span as a Tunable Hyperparameter:}  
The span $s$ thus acts as a \textit{data-preprocessing hyperparameter} that directly 
influences the effective information content of the auxiliary features. Unlike training 
parameters such as learning rate or batch size, the span alters the input representation 
itself, effectively redefining the train and test sets presented to the model. Consequently, 
span values cannot be arbitrarily fixed in advance; they must be carefully tuned.

To ensure unbiased evaluation, span selection is performed \textit{within the time-series 
cross-validation (CV) framework} used for model training. For each candidate span value, the 
smoothed feature series are re-estimated and the forecasting model is refitted on expanding 
windows of the training period. The optimal span is then chosen as the one minimizing average 
validation error across folds.

\subsubsection{\textbf{Stage 1 — Forecasting Yearly Features}}

The first stage of the proposed framework focuses on forecasting the auxiliary yearly features 
constructed in the previous subsection. Recall that for each district we derived yearly totals, 
monsoon totals, entropy, standard deviation, centroid, maximum, and quarterly proportions, 
resulting in nine feature series per district. These features provide complementary perspectives 
on the rainfall regime, but each is available only at yearly resolution, yielding  111 data 
points in total for the training period (1900--2010). Such short time series are inherently noisy and 
require parsimonious models that can leverage both temporal persistence and spatial dependence 
across districts.

To this end, we adopt a spatio-temporal regression strategy. Each yearly feature is modeled as a 
function of its own past values, lagged values from neighbouring districts, and a slow-moving 
trend component. This design captures both intra-district memory and inter-district correlation, 
while remaining computationally tractable for cross-validation and multi-step forecasting. The 
key components of this stage are detailed below.

\paragraph{\textbf{Regression formulation} }

For each smoothed yearly feature \(F_{d,t}\) (feature value for district \(d\) in year \(t\)) we fit a parsimonious spatio-temporal regression that combines (i) short own-series memory, (ii) lagged signals from a small set of geographically close neighbours, and (iii) compact short-run trajectory descriptors. The model is estimated separately for each district and each feature; parameters are selected by time-series cross-validation (expanding window) with LASSO regularization to avoid overfitting.

\medskip
\noindent\textbf{Model equation:} Let \(p\) be the number of own lags, \(k\) the number of neighbours and \(q\) the number of neighbour lags. Denote by \(n_j(d)\) the \(j\)-th nearest neighbour of district \(d\). The model  equation is
\begin{equation}\label{eq:stage1-main}
\begin{aligned}
{F}_{d,t} &= \beta_{0,d}
+ \sum_{\ell=1}^{p} \beta_{\ell,d}\, F_{d,t-\ell}
+ \sum_{j=1}^{k}\sum_{\ell=1}^{q} \gamma_{j,\ell,d}\, F_{\,n_j(d),\,t-\ell} \\
&\qquad + \delta_{1,d}\,\mathrm{Slope}_{d,t-1}
+ \delta_{2,d}\,\mathrm{MeanDiff}_{d,t-1} \\&\qquad
+ \delta_{3,d}\,\mathrm{Momentum}_{d,t-1}
+ \varepsilon_{d,t},
\end{aligned}
\end{equation}
where \(\varepsilon_{d,t}\) denotes an error term with zero mean. All predictors on the right-hand side are constructed using information available up to (and including) year \(t-1\). The description of the predictor $\mathrm{Slope}_{d,t-1}$ , $\mathrm{MeanDiff}_{d,t-1}$ and $\mathrm{Momentum}_{d,t-1}$ is as follows.

\medskip
\noindent\textbf{Short-run trajectory descriptors: } 
Let $\{F_{d,t}\}_{t=1}^T$ denote the smoothed yearly feature series for district $d$, 
where $F_{d,t}$ is the value at year $t$. To capture short-term dynamics, we use a 
look-back window of length $L$. At time $t$, the effective window size is 
\[
L' = \min(L,\,t-1),
\]  
so that the available values are 
\[
F_{d,t-L'},\,F_{d,t-L'+1},\,\dots,\,F_{d,t-1}
\]

From this window, three descriptors are constructed:

\begin{itemize}
  \item \textbf{Slope (trend):} The least-squares slope of the last $L'$ points:
\begin{equation}
  \mathrm{Slope}_{d,t-1} = 
  \frac{\sum_{j=1}^{L'} (j-\bar{j})(F_{d,t-L'+j-1}-\bar{F})}{\sum_{j=1}^{L'} (j-\bar{j})^2},
\end{equation}
  where 
  $\bar{j} = \tfrac{1}{L'}\sum_{j=1}^{L'} j$ 
  and 
  $\bar{F} = \tfrac{1}{L'}\sum_{j=1}^{L'} F_{d,t-L'+j-1}$

  \item \textbf{Mean difference (latest deviation):} The deviation of the most recent 
  value from its local mean:
\begin{equation}
  \mathrm{MeanDiff}_{d,t-1} = F_{d,t-1} - \frac{1}{L'} \sum_{j=1}^{L'} F_{d,t-L'+j-1}
\end{equation}

  \item \textbf{Momentum (directional consistency):} The fraction of upward moves in 
  the recent window:

{\footnotesize
\begin{equation}
  \mathrm{Momentum}_{d,t-1} = \frac{1}{L'-1} \sum_{j=1}^{L'-1} 
  \mathbf{1}\{F_{d,t-L'+j} - F_{d,t-L'+j-1} > 0\},
\end{equation}
}

  where $\mathbf{1}\{\cdot\}$ is the indicator function.
\end{itemize}

\medskip

These descriptors summarize different aspects of short-run behavior. The \textit{slope} 
captures the general trend in the recent past, showing whether the series has been 
moving upward or downward on average. The \textit{mean difference} compares the latest 
value with the recent average, highlighting whether the current year is unusually high 
or low relative to its short-term history. The \textit{momentum} reflects directional 
persistence: values close to 1 indicate mostly rising years, values near 0 indicate 
mostly falling years, and values around 0.5 suggest no consistent pattern. Together, 
these descriptors provide a compact characterization of local temporal dynamics.

These three descriptors provide low-dimensional summaries of recent behaviour (trend, anomaly, direction) and are especially valuable when long autoregressive lags are infeasible given the short yearly sample.

\paragraph{\textbf{Recursive Multi-step forecasting}}

For each district and each feature type, the smoothed yearly feature series depends on its own lagged values, the lagged values of neighbouring districts, and the short-run descriptors (slope, momentum, and mean difference). As these components are mutually dependent across districts, forecasting must be carried out jointly rather than independently. Accordingly, for any chosen parameter configuration, all yearly feature series are forecasted simultaneously in a recursive manner.

In the first forecast step, a one-year-ahead prediction is generated for every district and every feature. These forecasts are then used to recompute the short-run descriptors, which serve as inputs for the next prediction step. This process is repeated iteratively until the end of the forecasting horizon. If there are $D$ districts and $m$ feature types, then a total of $D \times m$ yearly time series are forecasted in parallel, where each feature set consists of 19 series (one per district). Once the full set of actual and forecasted yearly feature values is obtained for all districts, these are supplied as auxiliary inputs to the Stage~2 monthly forecasting model.

\paragraph{\textbf{ Parameters and parameter selection}}

The main hyperparameters for Stage~1 are collectively denoted by  (for feature set $i$ ) 
\[
\rho_i = (\text{span}_i,\ p_i,\ q_i,\ k_i,\ L_i, \lambda_i) ; i \in {1, ..., m}; m=9
\]
where $span_i$ controls the smoothing of yearly features, $p_i$ and $q_i$ denote the numbers of own and neighbour lags respectively, $k_i$ is the number of neighbouring districts considered, and $L_i$ is the window length used to compute the short-run descriptors (slope, momentum, and mean difference). In addition to these, an $\ell_1$ regularization parameter $\lambda_i$ is used in the regression model to prevent overfitting.

Since there are $m = 9$ feature types and $D = 19$ districts, allowing fully separate hyperparameters $\rho$ for every $(\text{district}, \text{feature})$ combination would lead to an extremely large search space. This would make randomized cross-validation computationally infeasible, as a prohibitively large number of samples would be required to explore the parameter space adequately.

To mitigate this, we impose a structural constraint: \textbf{for each feature type, a common parameter vector $\rho$ is used across all districts}. For example, for the feature ``yearly total'', a single span value is used to smooth all 19 yearly series. Likewise, all districts use the same values of $p$, $q$, and $k$, so that each district’s forecast depends on its past $p$ values, on the past $q$ values of its $k$ nearest neighbours (where neighbours are determined using district-level latitude--longitude coordinates), and on descriptors computed from the last $L$ observations. The LASSO penalty $\lambda$ is also kept common across districts.

A separate parameter vector $\rho$ is estimated independently for each feature type (e.g., yearly total, monsoon total, entropy, etc.). Although this constraint may incur a slight loss in flexibility compared to fully district-specific tuning, it substantially reduces the dimensionality of the hyperparameter space. As a result, the randomized search can cover a larger proportion of feasible configurations, improving both computational efficiency and robustness of the cross-validation procedure.

The hyperparameters are selected using time-series cross-validation. Although Stage~1 produces yearly feature forecasts, these are not evaluated independently, since their primary role is to provide meaningful signals—rather than precise predictions—for the Stage~2 MLP-based monthly model. In many cases, even if a yearly forecast is inaccurate in magnitude, it may still capture the correct direction or trend of the feature. For example, if the true yearly entropy increases during the validation period and the forecast also increases—albeit with error—this still supplies useful information to Stage~2. Penalizing such forecasts solely on Stage~1 RMSE would overlook their downstream utility.

Therefore, rather than optimizing Stage~1 and Stage~2 separately, we assess performance at the level of the full architecture. For a given Stage~1 parameter vector $\rho$, we generate yearly forecasts and pass them into Stage~2 under a given parameter configuration. The resulting monthly forecasts are then compared against observations over the validation period. This procedure is repeated across folds, and all hyperparameters from both stages are tuned jointly by minimizing the final monthly forecasting error. Thus, model selection is driven by end-to-end predictive accuracy rather than intermediate fit quality.

\begin{algorithm*}
\DontPrintSemicolon
\SetAlgoLined
\caption{HSTM: Joint Randomized Search with Expanding-Window TCV (avg-NRMSE)}
\label{alg:hstm}  
\KwIn{Monthly series $\{y_{d,t}\}$ for $d=1{:}D$, yearly feature sets $f=1{:}m$, Stage~1 hypergrid $\mathcal{H}_1$ over $\rho=(\text{span}_f,p_f,q_f,k_f,L_f,\lambda_f)$, Stage~2 hypergrid $\mathcal{H}_2$ over $\theta=(p_d,q_d,k_d,\text{units}_d,\eta_d,\alpha_d,\text{epochs}_d)$, folds $K$, validation length $v$, joint random samples $R$, seed $s$}
\KwOut{Selected hyperparameters $(\hat\rho,\hat\theta)$, fitted Stage~1 and Stage~2 models, holdout forecasts}
\SetKwFunction{BuildFolds}{BuildExpandingFolds}
\SetKwFunction{Smooth}{SmoothYearlyFeatures}
\SetKwFunction{FitSOne}{FitStageOne}
\SetKwFunction{FCstSOne}{ForecastStageOneRecursive}
\SetKwFunction{FitSTwo}{TrainStageTwoMLP}
\SetKwFunction{FCstSTwo}{ForecastMonthlyRecursive}
\SetKwFunction{NRMSE}{ComputeAvgNRMSE}
\SetKwComment{tcc}{\# }{}

\tcc{Initialize}
\textbf{set random seed} $s$\;
Construct yearly features $\{x^{(f)}_{d,T}\}$ from $\{y_{d,t}\}$ (Total, Monsoon, Entropy, SD, Centroid, Max, $Q_1,Q_2,Q_3$)\;
$F \leftarrow$ \BuildFolds{$K, v$} \tcp*{expanding-window folds over years}
$\text{bestScore} \leftarrow +\infty$, $(\hat\rho,\hat\theta)\leftarrow \varnothing$\;

\tcc{Joint randomized search over Stage~1 and Stage~2 spaces}
\For{$r \leftarrow 1$ \KwTo $R$}{
  Sample $\rho^{(r)} \sim \mathcal{H}_1$ \tcp*{per-feature: span shared across districts; $(p,q,k,L,\lambda)$}
  Sample $\theta^{(r)} \sim \mathcal{H}_2$ \tcp*{per-district: $(p,q,k,\text{units},\eta,\alpha,\text{epochs})$}
  $\text{CVscore} \leftarrow 0$\;
  
  \ForEach{fold $i \in F$}{
    \tcc{Stage~1 on training years; recursive yearly forecasts on validation years}
    \For{$f=1$ \KwTo $m$}{
      $\tilde{F}^{(f)}_{d,T} \leftarrow$ \Smooth{$x^{(f)}_{d,T}$, $\rho^{(r)}.\text{span}_f$} \tcp*{same span for all $d$ within feature $f$}
    }
    \FitSOne{$\tilde{F}^{(1{:}m)}_{d,T}$, $\rho^{(r)}$, fold $i$} \tcp*{LASSO with $(p,q,k,L,\lambda)$}
    $\widehat{F}^{(1{:}m)}_{d,T} \leftarrow$ \FCstSOne{$\tilde{F}^{(1{:}m)}_{d,T}$, $\rho^{(r)}$, fold $i$} \tcp*{joint, recursive across $D\times m$ series}
    
    \tcc{Stage~2: train MLP on training months; validate with Stage~1 predictions}
    \FitSTwo{$\{y_{d,t}\}$, $\theta^{(r)}$, fold $i$} \tcp*{use smoothed observed yearly features on train}
    $\widehat{Y}_{d,t} \leftarrow$ \FCstSTwo{$\{y_{d,t}\}$, $\widehat{F}^{(1{:}m)}_{d,T}$, $\theta^{(r)}$, fold $i$} \tcp*{recursive monthly forecasts}
    
    \tcc{Evaluate on fold $i$ using avg-NRMSE across districts}
    $\bar{\mathrm{NRMSE}}_i \leftarrow$ \NRMSE{$\widehat{Y}_{d,t}, Y_{d,t}$ on fold $i$}\;
    $\text{CVscore} \leftarrow \text{CVscore} + \bar{\mathrm{NRMSE}}_i$\;
  }
  $\text{CVscore} \leftarrow \text{CVscore}/K$ \tcp*{average over folds}
  \If{$\text{CVscore} < \text{bestScore}$}{
    $\text{bestScore} \leftarrow \text{CVscore}$; $(\hat\rho,\hat\theta) \leftarrow (\rho^{(r)},\theta^{(r)})$\;
  }
}

\tcc{Refit on full training with selected hyperparameters; produce holdout forecasts}
\For{$f=1$ \KwTo $m$}{
  $\tilde{F}^{(f)}_{d,T} \leftarrow$ \Smooth{$x^{(f)}_{d,T}$, $\hat\rho.\text{span}_f$}
}
\FitSOne{$\tilde{F}^{(1{:}m)}_{d,T}$, $\hat\rho$, full train}\;
$\widehat{F}^{(1{:}m)}_{d,T} \leftarrow$ \FCstSOne{$\tilde{F}^{(1{:}m)}_{d,T}$, $\hat\rho$, full train}\;
\FitSTwo{$\{y_{d,t}\}$, $\hat\theta$, full train}\;
$\widehat{Y}_{d,t}^{\text{holdout}} \leftarrow$ \FCstSTwo{$\{y_{d,t}\}$, $\widehat{F}^{(1{:}m)}_{d,T}$, $\hat\theta$, holdout}\;

\Return $(\hat\rho,\hat\theta)$, fitted Stage~1/Stage~2 models, $\widehat{Y}_{d,t}^{\text{holdout}}$
\end{algorithm*}

\subsubsection{\textbf{Stage 2: Monthly MLP Conditioned on Yearly Forecasts}}

In the second stage of the framework, the yearly forecasts obtained from Stage~1 are integrated 
into a monthly forecasting model. The aim is to combine high-frequency monthly dynamics with 
low-frequency yearly information to achieve robust long-horizon forecasts. The forecasting 
engine is a Multi-Layer Perceptron (MLP), which provides flexibility in capturing nonlinear 
dependencies while remaining computationally efficient. This section describes the input design, 
the network architecture and training procedure, and the correct handling of yearly features to 
avoid information leakage.

\paragraph{\textbf{Input Design}}
To predict monthly rainfall $y_{d,t}$ for district $d$ at month $t$, the input vector is 
constructed from the following components:

\begin{enumerate}
    \item \textit{Own monthly lags:}  
    Past observations of the same district,  
    \[
    y_{d,t-1},\; y_{d,t-2}, \dots, y_{d,t-p},
    \]  
    where $p$ is the number of monthly lags. These capture temporal persistence and 
    annual seasonality.

    \item \textit{Neighbouring monthly lags:}  
    For each of the $k_m$ nearest neighbours $n_1(d),\dots,n_{k_m}(d)$, include  
    \[
    y_{n_j(d),t-1},\dots,y_{n_j(d),t-q}, \quad j=1,\dots,k_m,
    \]  
    where $q$ is the neighbour lag depth. These encode spatial dependence in monthly rainfall.

    \item \textit{Auxiliary yearly features:}  
    For the year $T$ corresponding to month $t$, include the vector of Stage~1 predicted yearly 
    features,
    \[
    \widehat{\mathbf{G}}_{d,T} = 
    \big(\widehat{\text{Total}},\; \widehat{\text{Monsoon}},\; \widehat{\text{Entropy}},\; 
    \widehat{\text{SD}},\; \widehat{\text{Centroid}},\; \]
    \[\widehat{\text{Max}},\; 
    \widehat{Q}_1,\; \widehat{Q}_2,\; \widehat{Q}_3 \big).
    \]  
    These features summarize annual scale, distributional shape, variability, and intra-year 
    allocation of rainfall. The same yearly vector is repeated across all twelve months 
    of year $T$.

\end{enumerate}

The final input vector $\mathbf{x}_{d,t}$ concatenates monthly lags, neighbour lags,  and yearly 
features.

\paragraph{\textbf{Network Architecture and Training}}

The monthly forecasts are generated by an MLP of the form
\[
\hat{y}_{d,t} = f_\theta(\mathbf{x}_{d,t}),
\]
where $f_\theta(\cdot)$ is a feed-forward neural network with parameters $\theta$.

The neural network architecture consists of an input layer whose dimension equals the number of features in $\mathbf{x}_{d,t}$, followed by two hidden layers with 5--15 units each and ReLU activations, and a single linear output neuron for predicting monthly rainfall. L1 regularization is applied for weight decay and feature selection. The model is trained using the Adam optimizer with mean squared error (MSE) loss, for up to 200 epochs with early stopping based on validation RMSE. Input features are standardized using training-set statistics. The forecasting strategy is one-step ahead prediction, which is then applied recursively to generate multi-step forecasts.

\paragraph{\textbf{Parameters and parameter selection}}

 The overall structure of the Hierarchical Spatio-Temporal Model (HSTM) follows the benchmark Spatio-Temporal Lag Model (STLM), with the key enhancement that HSTM incorporates yearly aggregated features. These yearly features provide long-term contextual signals to the monthly forecasting model, helping to suppress short-term variability and improving stability over extended horizons. Consequently, the parameterization of HSTM closely mirrors that of STLM.

For each district $d$, the monthly rainfall series is forecasted using (i) its own $p_d$ past lags, (ii) the past $q_d$ lags of its $k_d$ nearest neighbouring districts, and (iii) the corresponding yearly feature values and their forecasts. These quantities constitute the \textit{structural parameters} of the model. In addition, the network includes \textit{training parameters}, namely the number of hidden units in the two-layer MLP, the learning rate $\eta_d$, the L1 regularization coefficient $\alpha_d$, and the number of training epochs. The full parameter vector for district $d$ is therefore defined as
\[
\theta_d = (p_d,\ q_d,\ k_d,\ \text{hidden layer units},\ \eta_d,\ \alpha_d,\ \text{epochs}).
\]
Collecting these across all districts yields the complete configuration
\[
\theta = (\theta_1,\ \theta_2,\ \dots,\ \theta_D),
\]
where $D$ denotes the total number of districts.

The hyperparameter selection procedure is analogous to that used in STLM and is conducted via time-series cross-validation. All district-level time series are forecasted jointly in a recursive manner to ensure proper propagation of spatial dependencies. Model performance is evaluated using the average Normalized Root Mean Squared Error (NRMSE) across districts, consistent with the STLM evaluation framework.


\begin{table}[h]
\centering
\caption{Final Stage~1 hyperparameters per yearly feature type.}
\label{tab:stage1-params}
\begin{tabular}{lcccccc}
\hline
\textbf{Feature Type} & Span & \textbf{$p$} & \textbf{$q$} & \textbf{$k$} & \textbf{$L$} & \textbf{$\lambda$} \\
\hline
Yearly Total            & 60  & 11 & 1 & 3 & 4 & $10^{-2}$ \\
Yearly Monsoon Total    & 40  & 2  & 4 & 1 & 3 & $5^{-2}$  \\
Yearly Entropy          & 30  & 13 & 3 & 1 & 3 & $5^{-3}$ \\
Yearly SD               & 50  & 8  & 3 & 3 & 4 & $5^{-2}$ \\
Yearly Centroid         & 40  & 14 & 4 & 3 & 3 & $10^{-1}$ \\
Yearly Max              & 30  & 7  & 5 & 2 & 2 & $5^{-2}$ \\
Yearly Q1 Proportion    & 20  & 5  & 2 & 3 & 2 & $10^{-2}$ \\
Yearly Q2 Proportion    & 50  & 8  & 3 & 2 & 2 & $5^{-3}$ \\
Yearly Q3 Proportion    & 50 & 10 & 1 & 3 & 5 & $10^{-2}$ \\
\hline
\end{tabular}
\end{table}

\begin{table*}[htbp]
\centering
\caption{Final Stage~2 hyperparameters per district (Randomized from defined search space)}
\label{tab:stage2-params-random}
\begin{tabular}{lcccccccc}
\hline
District & $p$ & $k$ & $q$ & Layers & Units & $\eta$ & $\alpha$ & Epochs \\
\hline
BANKURA         & 140 & 6 & 2 & 2 & (8,6)  & $10^{-3}$  & $10^{-3}$ & 80 \\
BIRBHUM         & 120 & 4 & 4 & 2 & (10,8) & $10^{-2}$  & $10^{-2}$ & 90 \\
BURDWAN         & 160 & 5 & 3 & 2 & (6,4)  & $10^{-4}$  & $10^{-4}$ & 50 \\
COOCH BEHAR     & 180 & 2 & 1 & 2 & (8,4)  & $10^{-3}$  & $10^{-1}$ & 70 \\
DARJEELING      & 100 & 5 & 4 & 2 & (4,2)  & $10^{-2}$  & $10^{-3}$ & 40 \\
HOOGLY          & 120 & 6 & 3 & 2 & (10,8) & $10^{-4}$  & $10^{-2}$ & 100 \\
HOWRAH          & 160 & 6 & 2 & 2 & (10,8) & $10^{-3}$  & $10^{-3}$ & 100 \\
JALPAIGURI      & 120 & 3 & 1 & 2 & (6,4)  & $10^{-2}$  & $10^{-2}$ & 60 \\
MALDA           & 100 & 7 & 2 & 2 & (4,2)  & $10^{-4}$  & $10^{-3}$ & 40 \\
MANBHUM PURULIA & 140 & 4 & 5 & 2 & (8,6)  & $10^{-2}$  & $10^{-1}$ & 80 \\
EAST MIDNAPORE  & 160 & 3 & 1 & 2 & (6,4)  & $10^{-3}$  & $10^{-2}$ & 60 \\
MURSHIDABAD     & 180 & 4 & 2 & 2 & (8,6)  & $10^{-4}$  & $10^{-4}$ & 70 \\
NADIA           & 100 & 3 & 4 & 2 & (10,8) & $10^{-2}$  & $10^{-1}$ & 90 \\
24 PARGANAS N   & 140 & 6 & 3 & 2 & (6,4)  & $10^{-3}$  & $10^{-3}$ & 60 \\
24 PARGANAS S   & 120 & 2 & 4 & 2 & (6,4)  & $10^{-4}$  & $10^{-2}$ & 50 \\
DINAJPUR NORTH  & 160 & 2 & 2 & 2 & (8,6)  & $10^{-2}$  & $10^{-1}$ & 70 \\
DINAJPUR SOUTH  & 140 & 2 & 4 & 2 & (4,2)  & $10^{-4}$  & $10^{-4}$ & 30 \\
WEST MIDNAPORE  & 160 & 3 & 1 & 2 & (2,2)  & $10^{-4}$  & $10^{-3}$ & 40 \\
KOLKATA         & 120 & 4 & 1 & 2 & (4,2)  & $10^{-3}$  & $10^{-3}$ & 50 \\
\hline
\end{tabular}
\end{table*}

\subsubsection{\textbf{Time-Series Cross-Validation and Parameter Tuning}}
The optimal parameter configuration is selected using time-series cross-validation with $K = 5$ folds. The cross-validation procedure and evaluation metric follow the setup described earlier in Subsubsection~\textit{Time-Series Cross-Validation and Parameter Tuning}. The performance metric is the average Normalized Root Mean Squared Error (NRMSE) across districts, averaged again over the $K$ folds. The best parameter configuration is defined as the one that minimizes $\bar{\mathrm{NRMSE}}$.

Given the high dimensionality of the parameter space across both stages of the model, random search is used instead of an exhaustive grid search.

\subsubsection*{Stage~1 (Yearly Feature Forecasting)}

For each feature type $i \in \{1,\dots,m\}$, with $m = 9$, a separate parameter vector is defined as
\[
\begin{aligned}
\text{span}_i &\in \{10, 30, \dots, 90\}, \\
p_i &\in \{1, 2, \dots, 15\}, \\
k_i &\in \{0, 1, 2, 3, 4, 5\}, \\
q_i &\in \{1, 2, 3, 4, 5\}, \\
L_i &\in \{3, 4, 5, 6, 7\}, \\
\lambda_i &\in \{10^{-4},\ 10^{-3},\ 5^{-3},\ 10^{-2},\ 5^{-2},\ 10^{-1}\}.
\end{aligned}
\]
Thus, each feature type has a parameter tuple
\begin{equation}
\rho_i = (\text{span}_i,\ p_i,\ q_i,\ k_i,\ L_i, \lambda_i),
\quad \text{for } i = 1,\dots,m,
\end{equation}
and the full Stage~1 configuration is
\begin{equation}
\rho = (\rho_1,\ \rho_2,\ \dots,\ \rho_m).
\end{equation}

\subsubsection*{Stage~2 (Monthly Forecasting)}

For each district $d \in \{1,\dots,D\}$, with $D = 19$, the Stage~2 MLP model has the following hyperparameters:
\[
\begin{aligned}
p_d &\in \{80, 100, 120, ..., 200\}, \\
q_d &\in \{1, 2, 3, ..., 10\}, \\
k_d &\in \{0, 1, 2, 3,...,  8\}, \\
\text{Hidden layers} &= 2, \\
\text{Units per layer}_d &\in \{2, 4, 6, 8, 10\}, \\
\eta_d &\in \{10^{-4},\ 10^{-3},\ 10^{-2}\}, \\
\alpha_d &\in \{10^{-4},\ 10^{-3},\ 10^{-2},\ 10^{-1}\}, \\
\text{Epochs}_d &\in \{20, 30, ..., 100\}.
\end{aligned}
\]
Hence, the Stage~2 parameter vector for district $d$ is
\begin{equation}
\theta_d = (p_d,\ q_d,\ k_d,\ \text{hidden layer units},\ \eta_d,\ \alpha_d,\ \text{epochs}),
\end{equation}
and the complete Stage~2 configuration is
\begin{equation}
\theta = (\theta_1,\ \theta_2,\ \dots,\ \theta_D).
\end{equation}

\subsubsection{Joint Random Search Over Both Stages}

In each iteration of the random search process, a candidate configuration $(\rho^\star,\ \theta^\star)$ is drawn. Under this configuration:

\begin{enumerate}
    \item Stage~1 is used to generate yearly forecasts.
    \item These forecasts are supplied to Stage~2 to produce monthly forecasts for all districts.
    \item For each fold $k$, the NRMSE is computed separately for each district and then averaged across districts to obtain $\bar{\mathrm{NRMSE}}_k$.
    \item The final score is obtained by averaging over folds:
    \[
    \bar{\mathrm{NRMSE}} = \frac{1}{K} \sum_{k=1}^{K} \bar{\mathrm{NRMSE}}_k.
    \]
\end{enumerate}

A total of 10{,}000 random configurations $(\rho^\star,\theta^\star)$ are evaluated, and the configuration with the lowest $\bar{\mathrm{NRMSE}}$ is chosen as the final model.

\subsubsection{Algorithmic Summary}

The pseudocode in Algorithm~\ref{alg:hstm} provides a structured overview of the full HSTM training procedure. It consolidates the two-stage architecture—Stage~1 yearly feature forecasting and Stage~2 monthly prediction—into a single nested cross-validation loop with joint hyperparameter tuning. Rather than optimizing the two stages independently, the algorithm evaluates each candidate configuration based on end-to-end forecasting performance, ensuring that the selected parameters maximise final monthly accuracy. This summary serves as an implementation blueprint, clarifying the flow of information between components and enabling reproducibility of the proposed framework.

\subsubsection{\textbf{Documentation}}

For reproducibility, we report the final chosen hyperparameters for both stages of the HSTM pipeline. 
Table~\ref{tab:stage1-params} lists the selected Stage~1 parameters for each yearly feature type, 
including the temporal and spatial lag orders $(p,q,k)$, the descriptor window length $L$, and the 
LASSO penalty $\lambda$. These settings define how slowly varying annual characteristics are propagated 
forward in time.

For completeness, Table~\ref{tab:stage2-params-random} summarizes the final Stage~2 (monthly MLP) hyperparameters 
used for district-level forecasting. Each district $d$ is associated with a configuration 
$\theta_d = (p_d, q_d, k_d, \text{units}, \eta_d, \alpha_d, \text{epochs})$. 
The values shown here are placeholders and will be updated with the final tuned settings.

\section{Performance Evaluation}

\begin{table}[h]
\centering
\caption{Baseline Model Performance (Naïve Forecast): District-wise sMAPE and NRMSE over 2011--2019.}
\label{tab:baseline-metrics}
\begin{tabular}{lcc}
\hline
\textbf{District} & \textbf{sMAPE (\%)} & \textbf{NRMSE} \\
\hline
BANKURA & 108.17 & 80.50 \\
BIRBHUM & 95.50 & 83.16 \\
BURDWAN & 106.04 & 85.11 \\
COOCH BEHAR & 100.39 & 77.94 \\
DARJEELING & 77.18 & 48.86 \\
HOOGLY & 96.84 & 72.17 \\
HOWRAH & 107.14 & 90.27 \\
JALPAIGURI & 69.61 & 53.49 \\
MALDA & 111.42 & 98.56 \\
MANBHUM PURULIA & 98.09 & 70.77 \\
EAST MIDNAPORE & 89.55 & 65.08 \\
MURSHIDABAD & 105.03 & 94.15 \\
NADIA & 101.77 & 88.46 \\
24 PARGANAS N & 103.21 & 90.49 \\
24 PARGANAS S & 99.12 & 69.88 \\
DINAJPUR NORTH & 93.92 & 95.48 \\
DINAJPUR SOUTH & 92.18 & 105.50 \\
WEST MIDNAPORE & 104.05 & 82.70 \\
KOLKATA & 98.55 & 68.17 \\
\hline
\end{tabular}
\end{table}

The goal of this section is to systematically assess and compare the forecasting performance of three competing models: (i) a naïve baseline forecast, (ii) the benchmark Spatio-Temporal Lag Model (STLM), and (iii) the proposed Hierarchical Spatio-Temporal Model (HSTM). All models are evaluated on a common holdout period spanning January~2011 to December~2019, ensuring a fair comparison under identical forecasting conditions. The holdout set covers nine full years of monthly rainfall observations for all districts, providing a sufficiently long horizon to assess both short-term accuracy and long-term error accumulation.

The results are organized in a structured manner: for each model we first report absolute error levels using district-wise performance tables and spatial error maps, followed by relative comparisons via percentage improvement plots. In addition to monthly-level accuracy, we further examine yearly aggregation behaviour through representative trajectories and sMAPE heatmaps to investigate temporal stability. Taken together, this section provides a comprehensive evaluation of the competing approaches in terms of accuracy, robustness, and spatial consistency.

\subsection{Evaluation Metrics}

The forecasting performance of all models is assessed using two complementary error measures: 
\textit{Normalized Root Mean Squared Error (NRMSE)} and 
\textit{Symmetric Mean Absolute Percentage Error (sMAPE)}. 
The NRMSE for district \(d\) over a forecast horizon of length \(T\) is defined as
\[
\mathrm{NRMSE}_d = 
\frac{\sqrt{\frac{1}{T}\sum_{t=1}^{T} (\hat{y}_{d,t} - y_{d,t})^2}}{\sigma_d},
\]
where \(y_{d,t}\) and \(\hat{y}_{d,t}\) denote the observed and forecasted rainfall values respectively, 
and \(\sigma_d\) is the standard deviation of the training-period data for district \(d\). 
Normalization by \(\sigma_d\) enables fair comparison across districts with different 
rainfall magnitudes.

\begin{figure}[h!]
    \centering
    \begin{tabular}{cc}
        \includegraphics[width=0.45\linewidth]{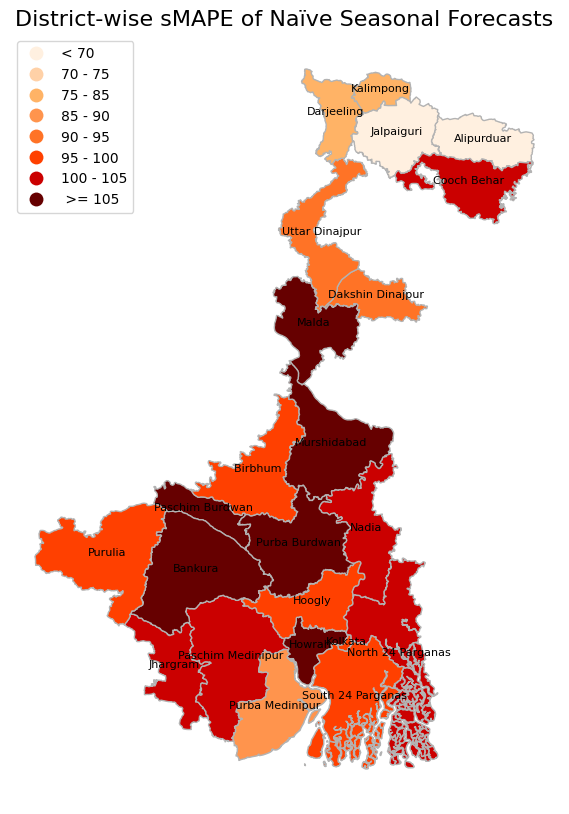} &
        \includegraphics[width=0.45\linewidth]{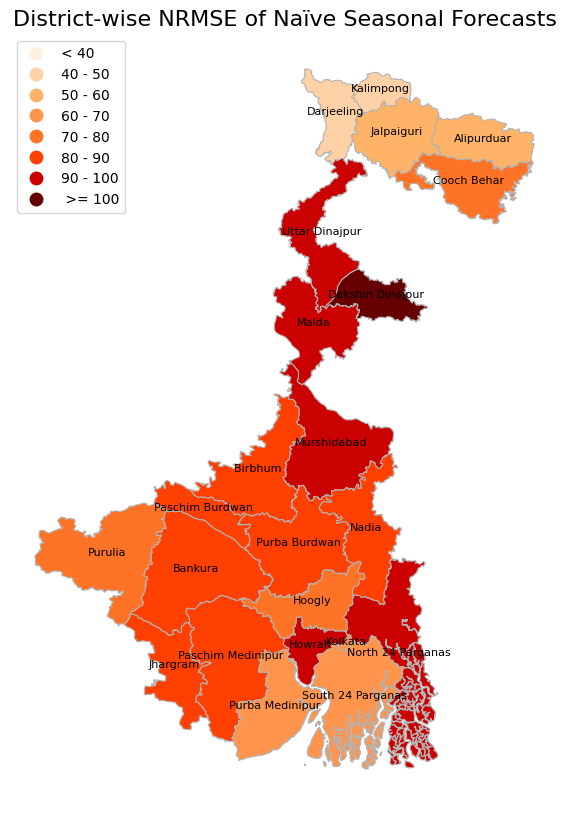} 

    \end{tabular}
    \caption{Spatial distribution of forecasting errors for the naïve baseline model over the holdout period (2011--2019). }
    \label{fig:naive_perf}
\end{figure}

The sMAPE is defined as
\begin{equation}
\mathrm{sMAPE}_d = 
\frac{100}{T} \sum_{t=1}^{T} 
\frac{|\hat{y}_{d,t} - y_{d,t}|}{(|\hat{y}_{d,t}| + |y_{d,t}|)/2}.
\end{equation}

Unlike the conventional Mean Absolute Percentage Error (MAPE), which becomes undefined 
or excessively inflated when \(y_{d,t} = 0\), sMAPE remains well-behaved even in dry months 
or for arid districts. Given that several districts exhibit zero or near-zero rainfall during 
parts of the year, sMAPE provides a more reliable and interpretable percentage-based error score. If for some month the actual and the predicted rainfall are both zero. We define the summand for that month to be zero as well in the formula for  the sMAPE calculation.

For each model, both metrics are computed \textit{per district} over the entire holdout period, and also at a \textit{year-wise resolution} for every district, as presented in the following subsections.

Together, NRMSE and sMAPE provide a balanced assessment of both \textit{scale-independent error}
(via NRMSE) and \textit{relative proportional error} (via sMAPE), ensuring robust evaluation 
across high-rainfall and low-rainfall regimes.

\subsection{Baseline Model Performance (Naïve Forecast)}

We begin the assessment by evaluating the naïve persistence-based baseline, which serves as the lower bound for all subsequent comparisons. The purpose of this subsection is to establish how much predictive skill can be achieved without any modeling sophistication, relying solely on past observations as direct forecasts. This enables a transparent quantification of the gains later provided by the STLM and HSTM architectures.

\subsubsection{Monthly Forecast Accuracy}

Table~\ref{tab:baseline-metrics} reports the district-wise forecasting error for the naïve baseline model over the holdout period (2011--2019), measured using sMAPE and NRMSE. As expected for a persistence-based benchmark, accuracy varies considerably across districts depending on the degree of temporal volatility.

\begin{figure}
    \centering
    \includegraphics[width=1\linewidth]{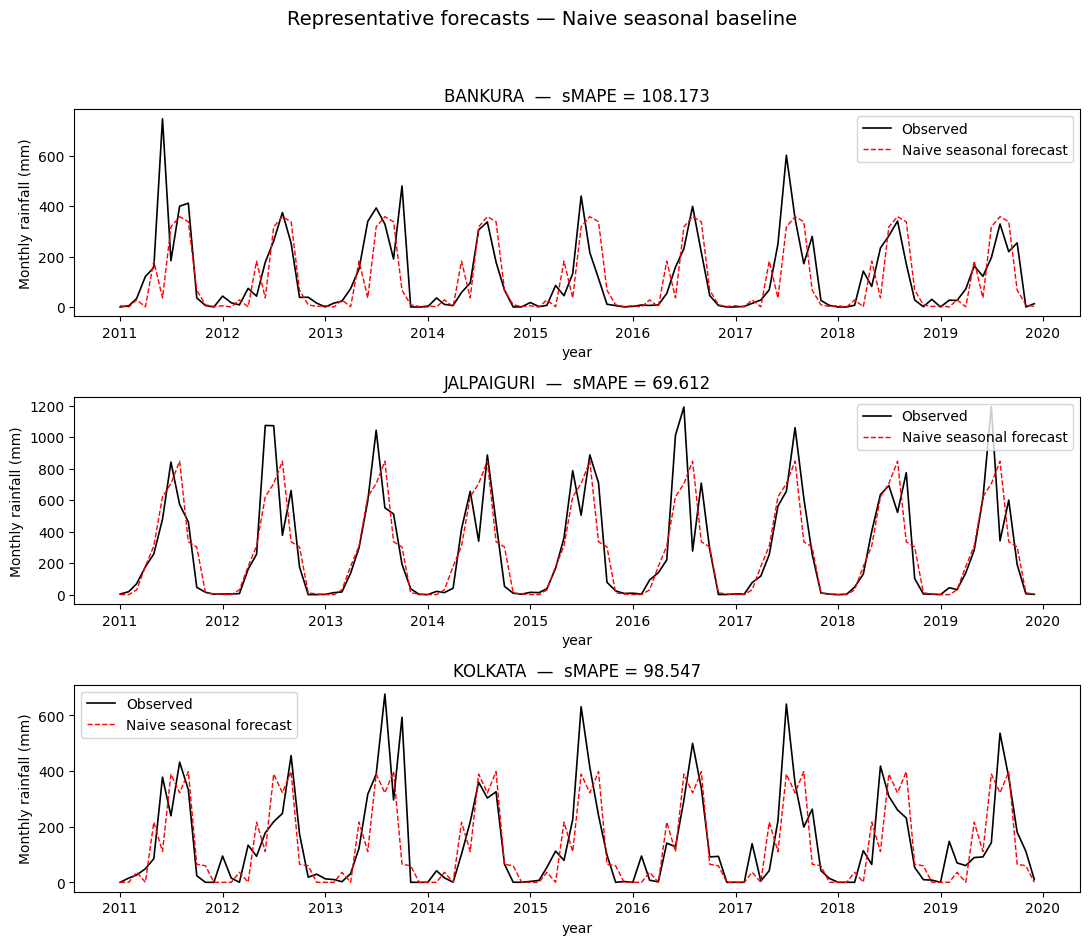}
    \caption{Representative baseline forecasts for three districts over the holdout period (2011--2019). 
    Observed rainfall (solid black) is compared against naïve seasonal forecasts (dashed red). 
    Jalpaiguri exhibits relatively stable seasonal alignment, whereas Bankura and Kolkata display substantial deviations in monsoon peak magnitude.}

    \label{fig:naive_act_f}
\end{figure}

\begin{figure}
    \centering
    \includegraphics[width=1\linewidth]{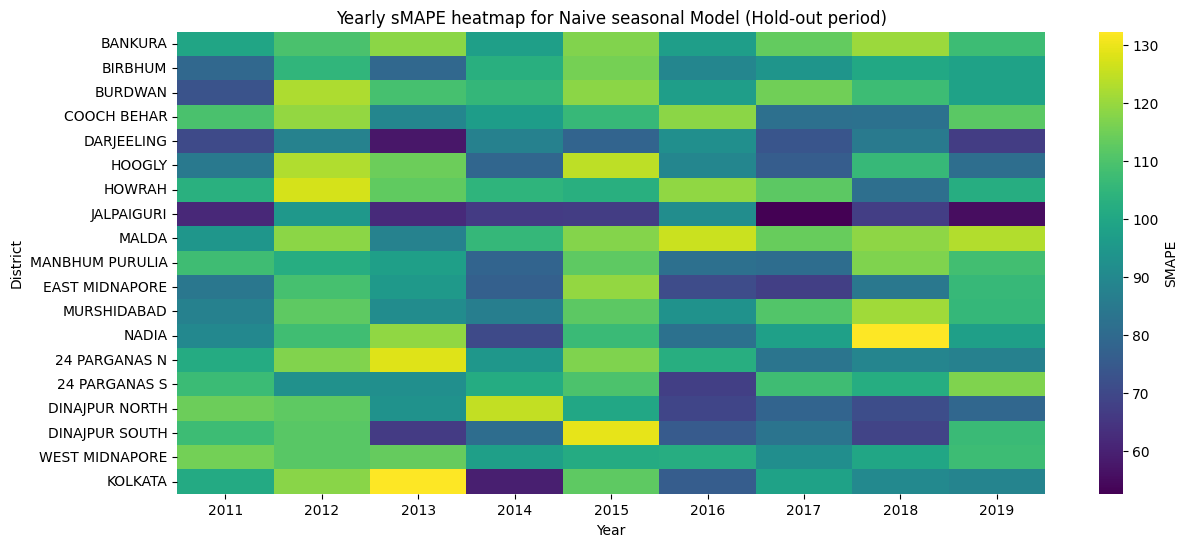}
    \caption{Year-wise sMAPE of naïve baseline forecasts across all districts during the 2011--2019 holdout period. 
    Each cell represents the annual sMAPE for a given district-year pair, with lighter shades indicating higher error. 
    }

    \label{fig:naive_ht}
\end{figure}

Districts such as \textit{Jalpaiguri} and \textit{Darjeeling}, which exhibit relatively smooth seasonal patterns, achieve comparatively lower sMAPE values (69.6\% and 77.2\%, respectively). In contrast, highly variable regions such as \textit{Malda} and \textit{Bankura} produce sMAPE values above 110\%, indicating substantial deviation between forecasted and actual rainfall. A similar pattern is reflected in the NRMSE values, with the best-performing regions showing errors around 50--60 units while the worst exceed 90 units.

\begin{table}[h]
\centering
\caption{Benchmark Model Performance (STLM): District-wise sMAPE and NRMSE over 2011--2019.}
\label{tab:stlm-metrics}
\begin{tabular}{lcc}
\hline
\textbf{District} & \textbf{sMAPE (\%)} & \textbf{NRMSE} \\
\hline
BANKURA & 89.92 & 71.32 \\
BIRBHUM & 87.54 & 75.23 \\
BURDWAN & 80.53 & 73.02 \\
COOCH BEHAR & 81.90 & 61.28 \\
DARJEELING & 57.67 & 35.03 \\
HOOGLY & 83.82 & 56.89 \\
HOWRAH & 86.71 & 74.74 \\
JALPAIGURI & 68.68 & 42.30 \\
MALDA & 116.52 & 101.92 \\
MANBHUM PURULIA & 89.38 & 65.15 \\
EAST MIDNAPORE & 90.33 & 67.49 \\
MURSHIDABAD & 81.86 & 75.86 \\
NADIA & 85.35 & 76.80 \\
24 PARGANAS N & 81.76 & 57.74 \\
24 PARGANAS S & 88.05 & 82.25 \\
DINAJPUR NORTH & 78.79 & 77.93 \\
DINAJPUR SOUTH & 115.35 & 91.66 \\
WEST MIDNAPORE & 79.99 & 74.42 \\
KOLKATA & 88.74 & 75.06 \\
\hline
\end{tabular}
\end{table}

Overall, the baseline model provides a conservative benchmark but struggles to capture interannual variability and spatial heterogeneity, especially in districts with irregular or intermittent rainfall behaviour. These results establish a lower bound against which the more sophisticated STLM and HSTM models will be evaluated.

\subsubsection{Spatial Error Distribution}

Figure~\ref{fig:naive_perf} illustrate the spatial distribution of forecasting errors across districts for the naïve baseline model, visualized using sMAPE and NRMSE. Both the maps in the Figure~\ref{fig:naive_perf} reveal clear geographical patterns in predictability.

Hill-dominated northern districts such as \textit{Darjeeling} and \textit{Jalpaiguri} exhibit comparatively lower error levels, suggesting that their seasonal rainfall patterns are more regular and easier to extrapolate using simple persistence. In contrast, western and south-central districts such as \textit{Bankura}, \textit{Malda}, and \textit{Howrah} display consistently high error magnitudes under both metrics, indicating substantial interannual variability that cannot be captured by static baselines.

The close agreement between the sMAPE and NRMSE maps confirms that the baseline model struggles in the same regions across both absolute and relative error metrics. These spatial discrepancies highlight the need for more expressive models capable of leveraging both temporal memory and spatial connectivity, which motivates the use of STLM and HSTM in subsequent sections.

\subsubsection{Yearly Forecast Consistency}

To further examine the temporal stability of the baseline model, Figure~\ref{fig:naive_act_f} presents forecast trajectories for three representative districts: Bankura (high-error), Jalpaiguri (low-error), and Kolkata (moderate-error). The naïve forecast reproduces the broad seasonal patterns but fails to capture interannual fluctuations in peak magnitude, leading to systematic over- or under-estimation in extreme rainfall years. This behaviour is especially evident in Bankura, where sharp monsoon spikes are consistently misaligned in both timing and intensity.


\begin{figure}[h!]
    \centering
    \begin{tabular}{cc}
        \includegraphics[width=0.465\linewidth]{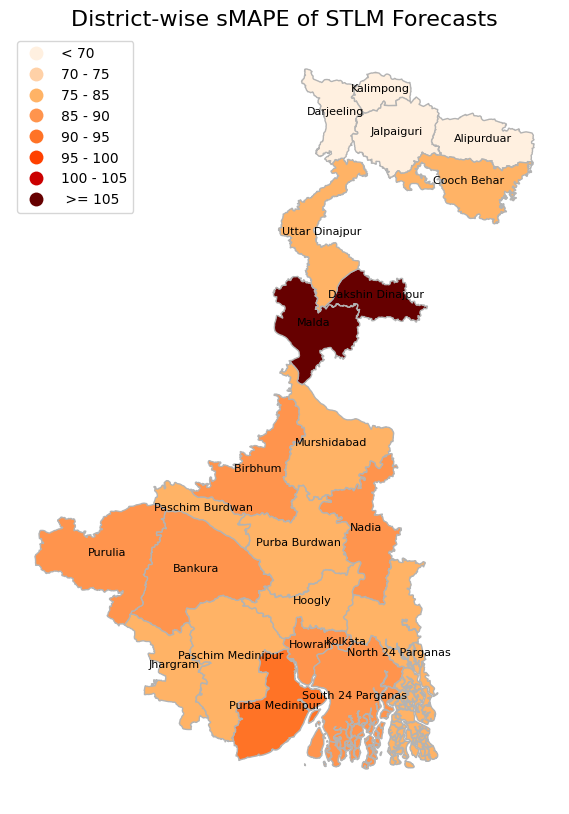} &
        \includegraphics[width=0.47\linewidth]{images/naive_nrmse_1.png} 

    \end{tabular}
    \caption{Spatial distribution of forecasting errors for STLM over the holdout period (2011--2019). }
    \label{fig:stlm_perf}
\end{figure}

\begin{figure}[h!]
    \centering
    \begin{tabular}{cc}
        \includegraphics[width=0.465\linewidth]{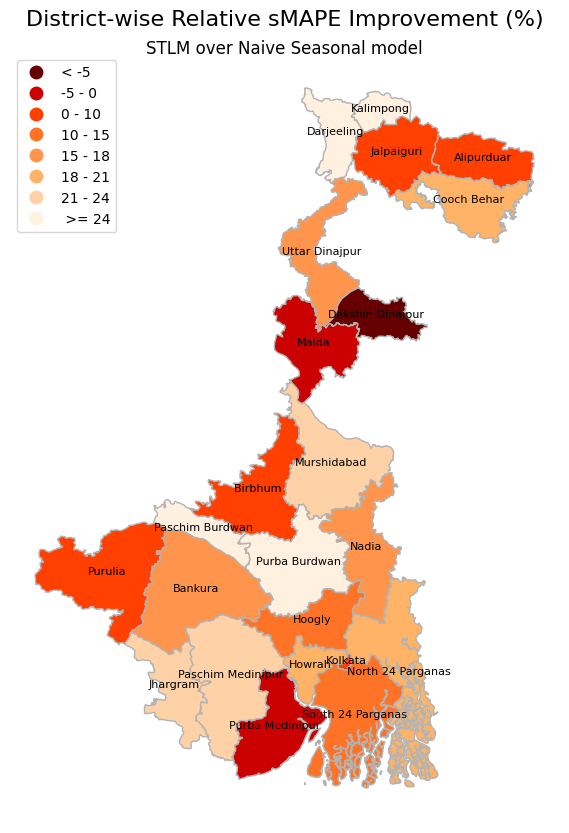} &
        \includegraphics[width=0.47\linewidth]{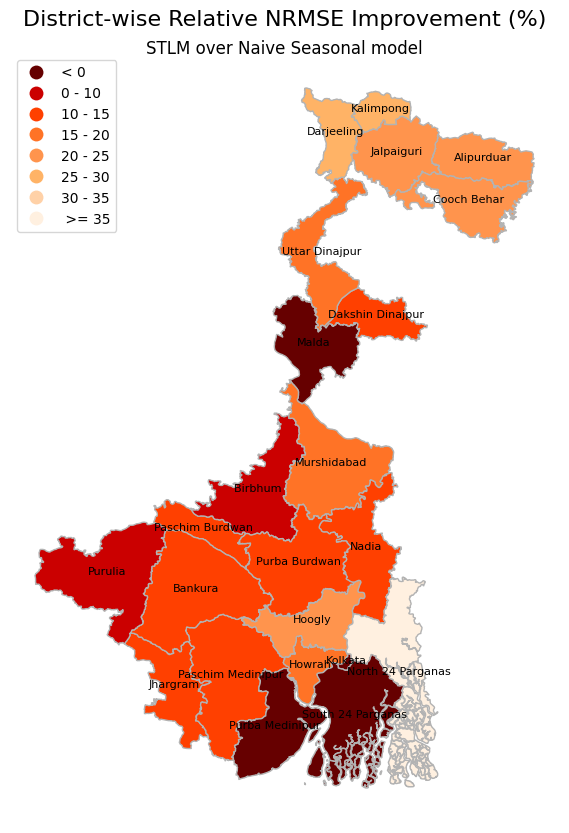} 

    \end{tabular}
    \caption{Spatial distribution of forecasting errors for STLM over the holdout period (2011--2019). }
    \label{fig:stlm_naive_perf}
\end{figure}


A more comprehensive view of temporal error propagation is provided in the sMAPE heatmap shown in Figure~\ref{fig:naive_ht}. Each row corresponds to a district and each column to a year within the 2011--2019 holdout period. The heatmap reveals that several districts exhibit large year-to-year variability in forecast accuracy, with some years (e.g., 2012 and 2016) emerging as uniformly challenging across most regions. Conversely, districts such as Jalpaiguri and Darjeeling maintain relatively stable sMAPE values across years, reinforcing the earlier observation that persistence-based models perform best in regions with regular seasonal structure. These patterns highlight the limitations of naïve extrapolation in capturing nonstationary rainfall behaviour and motivate the need for models with stronger temporal learning capacity.

\subsection{Benchmark Model Performance (STLM)}

\begin{figure}
    \centering
    \includegraphics[width=1\linewidth]{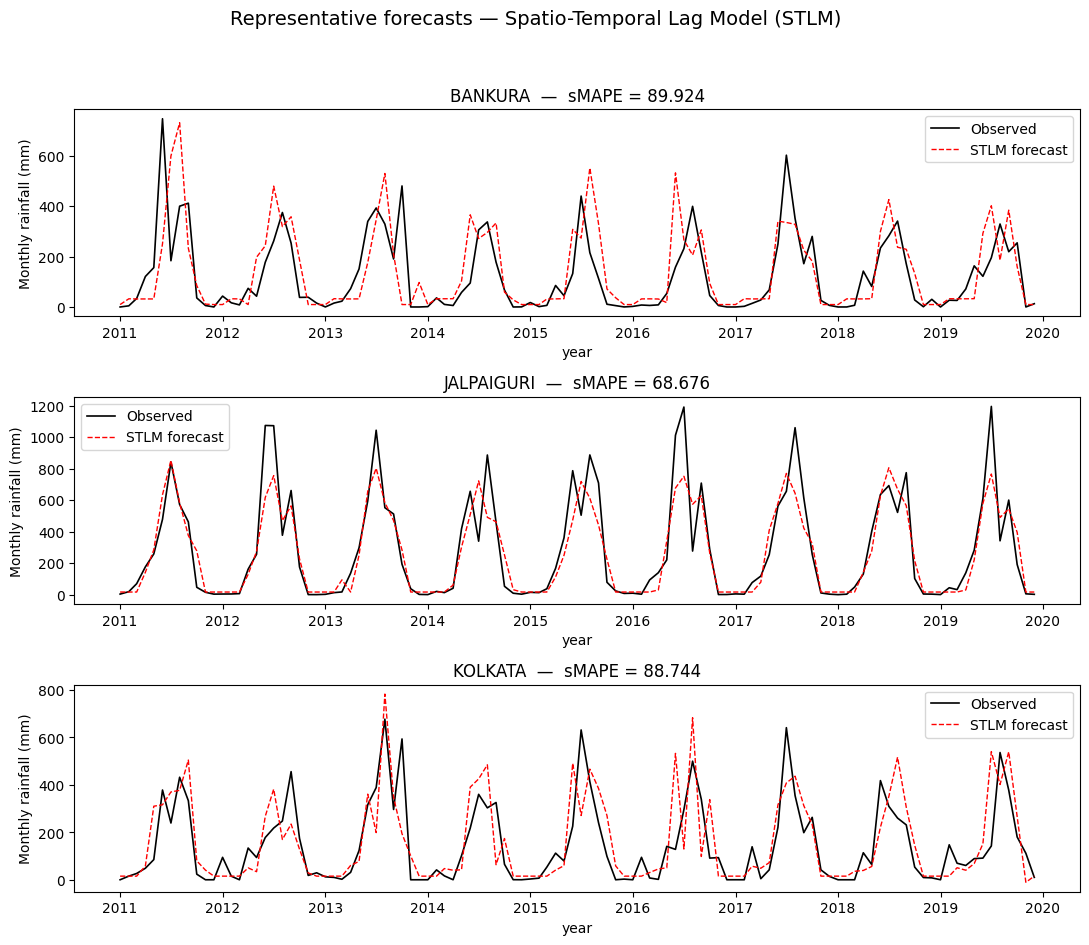}
    \caption{Representative yearly forecast trajectories under the STLM benchmark for three districts over the 2011--2019 hold-out period. 
The model captures seasonal recurrence reasonably well, but deviations in peak magnitude and phase remain visible in more volatile districts such as Bankura and Kolkata.}

    \label{fig:stlm_model_act_f}
\end{figure}

\begin{figure}
    \centering
    \includegraphics[width=1\linewidth]{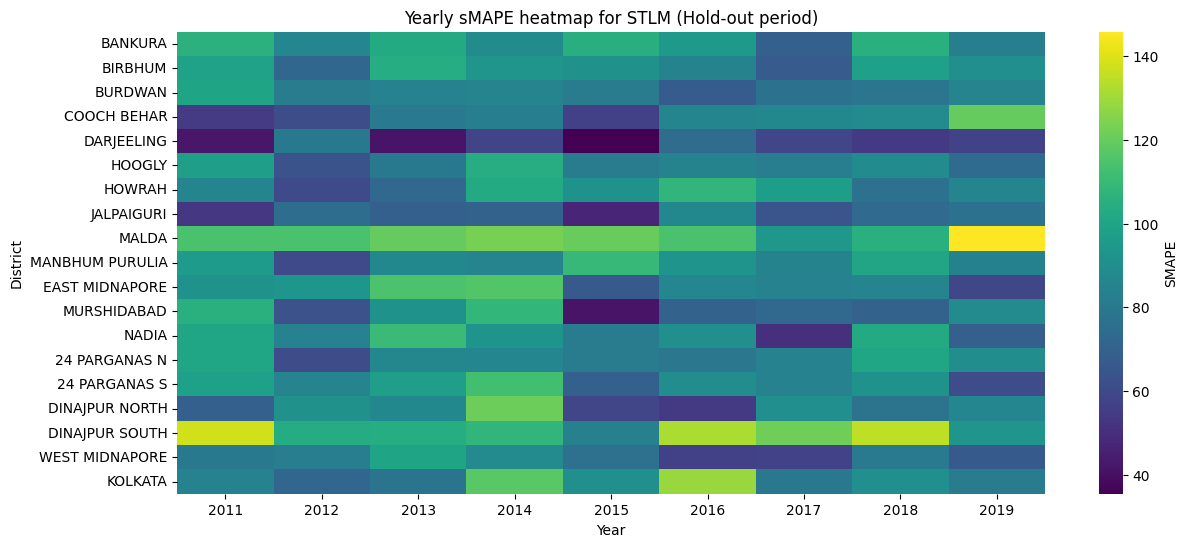}
    \caption{Year-wise sMAPE distribution of STLM forecasts across districts for the 2011--2019 hold-out period. 
Darker bands indicate improved temporal stability compared to the naïve baseline, though error spikes persist in highly erratic regions such as Malda and South Dinajpur.}

    \label{fig:stlm_model_ht}
\end{figure}

\begin{table}[h]
\centering
\caption{District-wise sMAPE and NRMSE of HSTM forecasts (Hold-out: 2011--2019)}
\label{tab:hstm_metrics}
\begin{tabular}{lcc}
\hline
\textbf{District} & \textbf{sMAPE (\%)} & \textbf{NRMSE} \\
\hline
BANKURA & 77.19 & 62.10 \\
BIRBHUM & 77.44 & 56.55 \\
BURDWAN & 70.56 & 50.77 \\
COOCH BEHAR & 70.90 & 54.64 \\
DARJEELING & 53.68 & 29.93 \\
HOOGLY & 75.36 & 49.66 \\
HOWRAH & 82.33 & 64.21 \\
JALPAIGURI & 64.24 & 36.75 \\
MALDA & 89.13 & 68.83 \\
MANBHUM PURULIA & 81.74 & 54.91 \\
EAST MIDNAPORE & 75.31 & 49.98 \\
MURSHIDABAD & 76.68 & 61.98 \\
NADIA & 76.95 & 58.33 \\
24 PARGANAS N & 72.59 & 49.61 \\
24 PARGANAS S & 79.12 & 54.63 \\
DINAJPUR NORTH & 85.39 & 56.47 \\
DINAJPUR SOUTH & 97.82 & 85.93 \\
WEST MIDNAPORE & 71.14 & 57.57 \\
KOLKATA & 77.97 & 51.40 \\
\hline
\end{tabular}
\end{table}

\subsubsection{Monthly Forecast Accuracy}

Table~\ref{tab:stlm-metrics} presents the district-wise sMAPE and NRMSE for the STLM benchmark model over the holdout period (2011--2019). Compared to the naïve baseline, the STLM achieves a consistent reduction in both error metrics across most districts. For instance, in Bankura the sMAPE drops from 108.2\% under the baseline to 89.9\%. The improvement is more pronounced in districts with regular rainfall dynamics, such as Darjeeling, where sMAPE decreases from 77.2\% to 57.7\%, and NRMSE falls from 48.9 to 35.0.

\begin{figure}[h]
\centering
\includegraphics[width=0.48\linewidth]{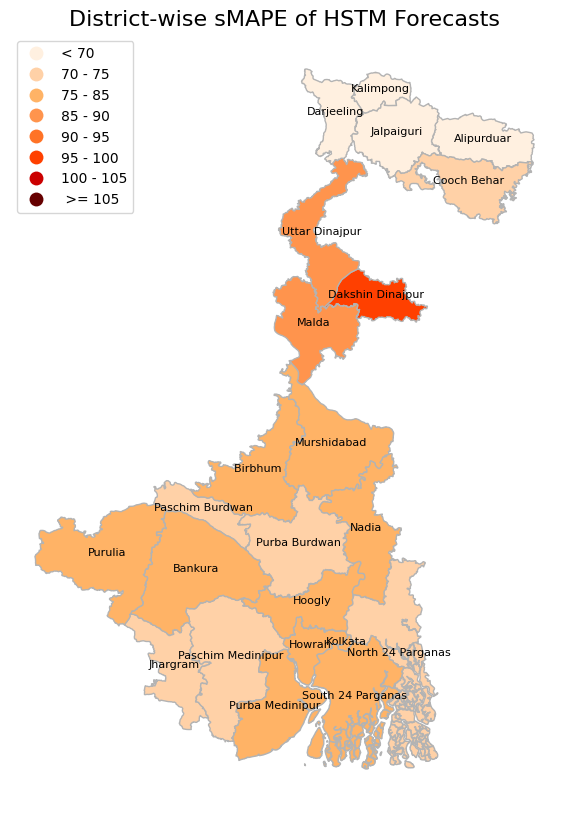}
\includegraphics[width=0.48\linewidth]{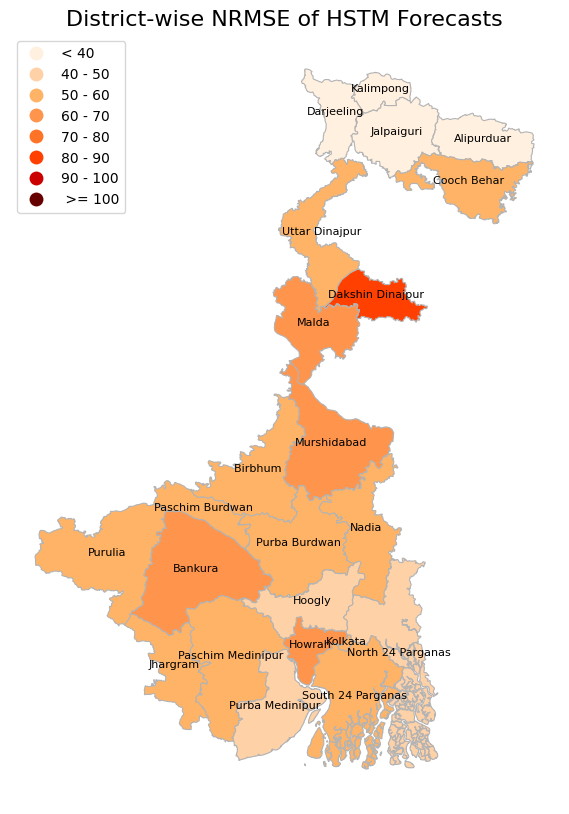}
\caption{District-wise sMAPE (left) and NRMSE (right) of HSTM forecasts.}
\label{fig:hstm_spatial}
\end{figure}

However, STLM still struggles in highly erratic regions such as Malda and South Dinajpur, where sMAPE exceeds 115\% and NRMSE remains above 90, indicating that  lag-based propagation alone cannot fully capture extreme interannual fluctuations. Overall, while STLM substantially improves upon naïve persistence, especially in climatically stable districts, it exhibits limitations in noise-prone or highly nonstationary locations. These residual discrepancies motivate the incorporation of higher-level temporal information in the proposed HSTM architecture.

\subsubsection{Spatial Error Distribution}

Figure~\ref{fig:stlm_perf}  provides a spatial overview of forecasting accuracy across districts. The error hot-spots align with those observed in the baseline, although the overall intensity is lower. Hill districts such as Darjeeling and Jalpaiguri exhibit the lowest error levels, reflecting their predictable monsoon structure. In contrast, central and western districts such as Malda, South Dinajpur, and 24 Parganas South continue to display high sMAPE and NRMSE values. These spatial trends confirm that the STLM captures local temporal autocorrelation effectively but lacks mechanism to account for broader-scale shifts or evolving annual patterns --- a gap addressed explicitly by the HSTM model in the next section.

\subsubsection{Improvement over Baseline}

The performance gains of STLM relative to the naïve persistence model are illustrated in 
Figure~\ref{fig:stlm_naive_perf} and Table~\ref{tab:stlm_naive} (refer the Appendix). The majority of districts exhibit
substantial reductions in both sMAPE and NRMSE, confirming that the incorporation of
spatio-temporal lags provides meaningful predictive benefit over simple extrapolation.

The strongest improvements are observed in districts with moderately regular rainfall
patterns. For example, Darjeeling and West Midnapore record sMAPE reductions exceeding
25\%, accompanied by over 28\% and 10\% reductions in NRMSE, respectively. Similarly,
Burdwan, Murshidabad, and 24 Parganas North achieve more than 20\% improvement in at
least one of the error measures. These systematic gains highlight the ability of STLM
to leverage both temporal persistence and spatial transferability.

\begin{figure}[ht]
    \centering
    \includegraphics[width=0.48\linewidth]{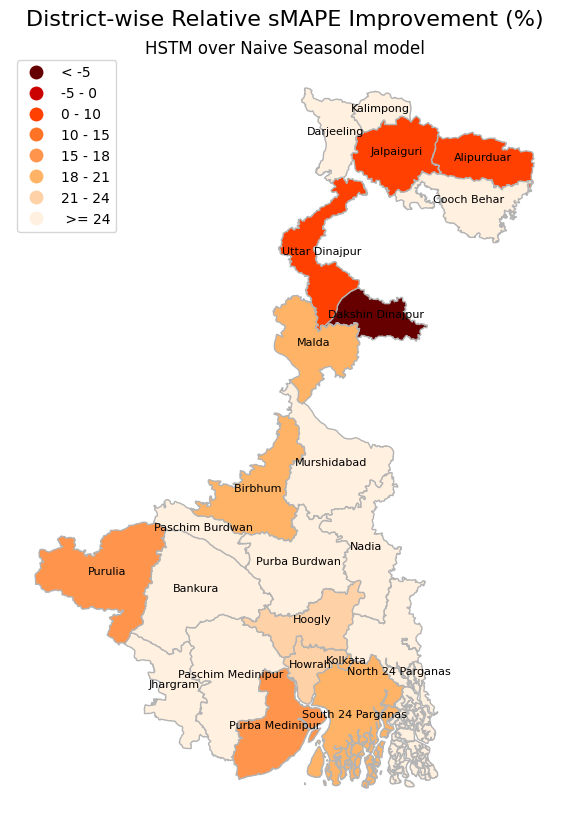}
    \includegraphics[width=0.48\linewidth]{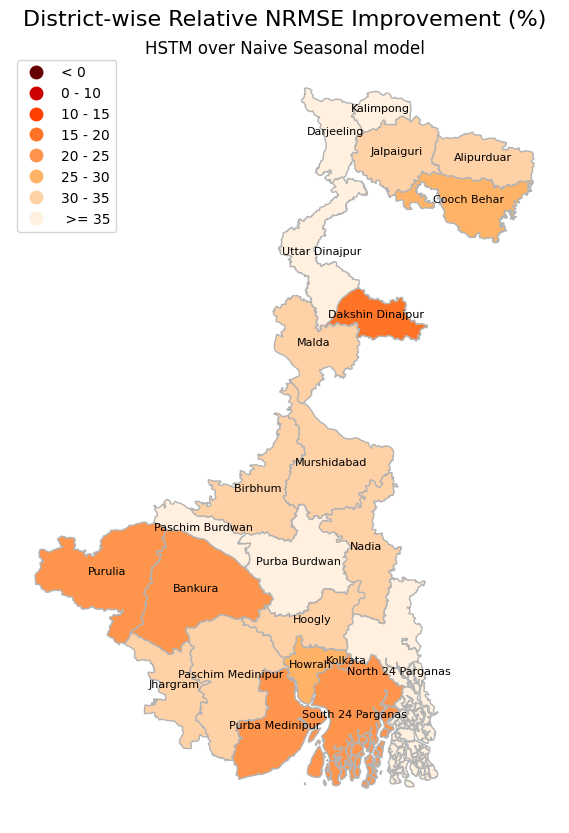}
    \caption{Spatial distribution of relative improvement of HSTM over the naïve baseline in terms of (left) sMAPE and (right) NRMSE.}
    \label{fig:imp_baseline_maps}
\end{figure}

However, improvement is not uniform across the region. Volatile or highly
non-stationary districts such as Malda, South Dinajpur, and parts of East Midnapore
display marginal or even negative change in sMAPE, despite achieving modest gains in
NRMSE. This suggests that while STLM better aligns magnitude on average, it may still
misrepresent phase or seasonal timing in certain locations.

Overall, STLM significantly outperforms the baseline in most areas, but its
limitations in capturing complex distributional shifts motivate the more flexible
hierarchical structure introduced in HSTM.

\subsubsection{Yearly Forecast Consistency}

Figure~\ref{fig:stlm_model_act_f} illustrates representative forecast trajectories for three districts under STLM. 
Compared to the naïve baseline (shown earlier in Figure~\ref{fig:naive_act_f}), 
the STLM is able to better align with both the timing and magnitude of seasonal peaks, particularly in districts such as Jalpaiguri. 
While the naïve model simply repeats the previous year's pattern, leading to rigid phase-locked oscillations, 
the STLM adapts to interannual fluctuations through the inclusion of lagged neighbour effects. 
This yields visibly smoother residuals and reduced over/undershooting of peak monsoon intensity.
However, in highly irregular regions such as Bankura and Kolkata, the STLM still exhibits occasional lag in responding to abrupt anomalies, 
suggesting that purely linear lag propagation is insufficient for extreme-event adaptation.

The sMAPE heatmap in Figure~\ref{fig:stlm_model_ht} provides a year-by-year error profile across districts. 
Relative to the baseline heatmap (Figure~\ref{fig:naive_ht}), the STLM version displays two favourable trends: 
(i) a general darkening of colour across most districts, indicating reduced error magnitudes, 
and (ii) improved temporal stability, with fewer sharp year-level error spikes. 
Notably, previously volatile districts such as Murshidabad and West Midnapore show consistent improvement over multiple years. 
However, the gains are not universal—districts like Malda still retain pockets of high sMAPE even under STLM, 
underscoring the need for additional long-term context, which the HSTM model provides by incorporating yearly-scale signals.

Overall, STLM improves forecast adaptability compared to the naïve model, but 
its year-wise residual patterns reveal that certain structural deviations cannot be corrected through lag-based learning alone. 
This motivates the transition to the hierarchical HSTM framework.

\subsection{Proposed Model Performance (HSTM)}

\begin{figure}[ht]
    \centering
    \includegraphics[width=0.48\linewidth]{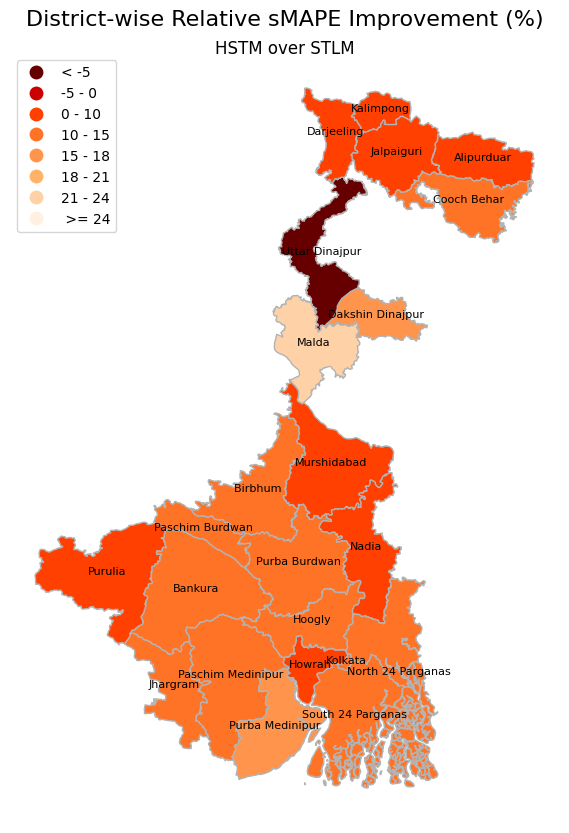}
    \includegraphics[width=0.48\linewidth]{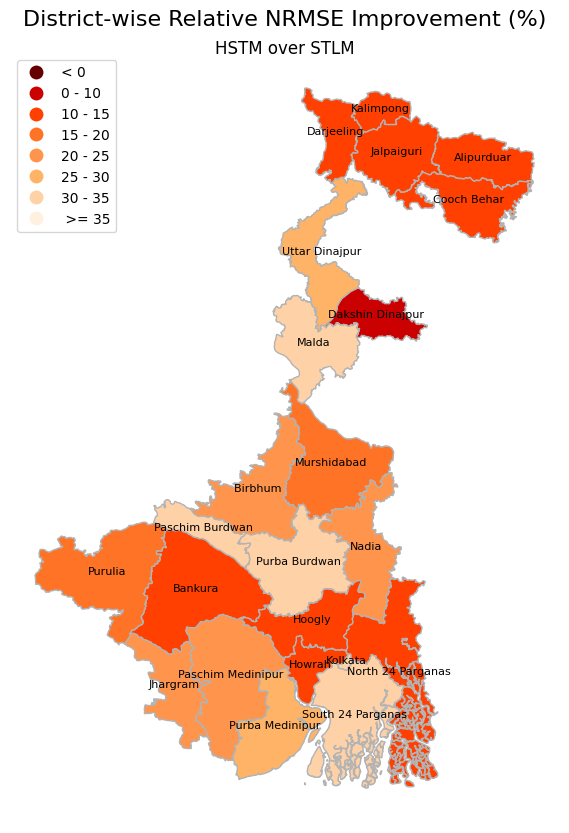}
    \caption{Spatial distribution of relative improvement of HSTM over the STLM benchmark in terms of (left) sMAPE and (right) NRMSE.}
    \label{fig:imp_stlm_maps}
\end{figure}

We now evaluate the forecasting accuracy of the proposed Hierarchical Spatio-Temporal Model (HSTM) on the 2011--2019 hold-out period. Unlike the naïve baseline and the purely autoregressive STLM benchmark, HSTM integrates both high-frequency monthly dynamics and low-frequency annual signals through a two-stage architecture. The goal of this subsection is to assess whether this hierarchical conditioning leads to tangible improvements in predictive stability and error reduction across districts. We begin with an overall assessment of monthly forecast accuracy, followed by a spatial analysis of residual patterns to understand the geographical strengths and limitations of the model.

\subsubsection{Monthly Forecast Accuracy}

Table~\ref{tab:hstm_metrics} reports the district-wise sMAPE and NRMSE scores for the proposed HSTM model on the 2011--2019 hold-out period. Overall, the model demonstrates substantially lower forecasting errors compared to the naïve baseline and achieves competitive accuracy across both dry and wet districts. 

The majority of districts record \textbf{sMAPE values between 70\% and 80\%}, indicating that the model captures seasonal dynamics while mitigating large proportional deviations. In terms of NRMSE, \textbf{errors are largely contained within the 50--65 range}, except for inherently high-variance locations such as Malda and Dinajpur South, which remain challenging due to extreme rainfall fluctuations.

Notably, \textbf{Darjeeling, Jalpaiguri, and Cooch Behar}---districts influenced by orographic rainfall mechanisms---show \textbf{sMAPE below 65\%} and low NRMSE values, suggesting that the hierarchical fusion of yearly features with spatial lags is particularly effective when temporal persistence and spatial coherence are strong.

\subsubsection{Spatial Error Distribution}

Figure~\ref{fig:hstm_spatial} visualizes the spatial distribution of forecast errors across districts. Both \textbf{sMAPE and NRMSE exhibit coherent geographical patterns}, with \textbf{central and southern districts} (e.g., Burdwan, Bankura, East Midnapore) showing \textbf{consistently lower errors}, whereas \textbf{northern regions}---notably Malda and Dinajpur South---remain comparatively difficult to predict.

\begin{figure}
    \centering
    \includegraphics[width=1\linewidth]{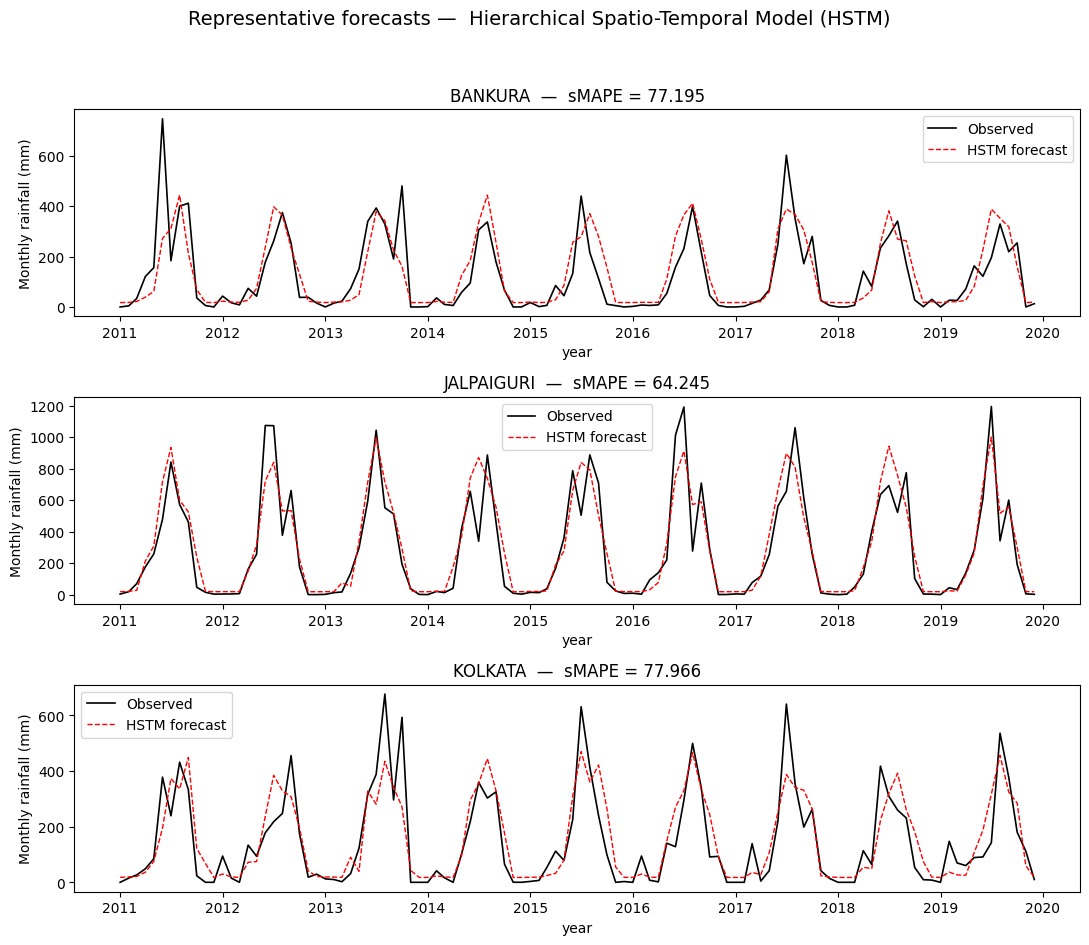}
    \caption{Observed versus forecasted monthly rainfall for three representative districts (Bankura, Jalpaiguri, Kolkata) under the proposed HSTM model. The forecasts closely track observed seasonality and interannual variability, with substantially reduced over/under-shoots compared to the Naïve and STLM models.}

    \label{fig:hstm_model_act_f}
\end{figure}

\begin{figure}
    \centering
    \includegraphics[width=1\linewidth]{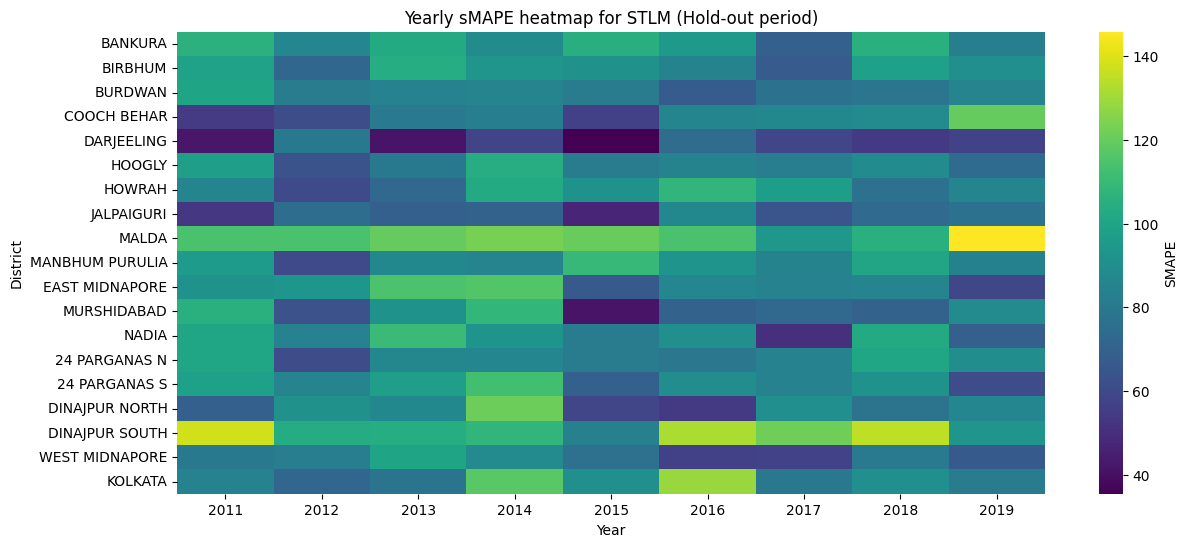}
    \caption{Year-wise sMAPE heatmap for HSTM forecasts across all districts during the hold-out period (2011--2019). Compared to the Naïve and STLM models, HSTM exhibits lower year-to-year volatility and fewer instances of extreme error spikes, particularly in high-variability districts such as Darjeeling and Burdwan.}

    \label{fig:hstm_model_ht}
\end{figure}

\subsubsection{Improvement over Baseline and STLM}

To quantify the relative gains achieved by the proposed HSTM framework, we compute the percentage reduction in both sMAPE and NRMSE with respect to (i) the naïve seasonal baseline and (ii) the STLM benchmark. Table~\ref{tab:hstm_naive} and Table~\ref{tab:hstm_stlm} (reported in Appendix~X for brevity) provide the district-wise improvement values, while Figures~\ref{fig:imp_baseline_maps} and~\ref{fig:imp_stlm_maps} summarize the same information spatially via choropleth maps.

\medskip
\noindent\textbf{Improvement over the naïve baseline.}
HSTM consistently outperforms the baseline across nearly all districts, with typical reductions of 20--35\% in NRMSE and 15--30\% in sMAPE. The gains are particularly pronounced in coastal and western districts such as \textit{Paschim Midnapore}, \textit{Bankura}, and \textit{Murshidabad}, where HSTM reduces error by more than 30\% in both metrics. Only a few districts---notably \textit{North \& South Dinajpur},  \textit{Jalpaiguri}---exhibit marginal or negative improvement in sMAPE, although even in these cases the NRMSE reduction remains positive. This indicates that while amplitude matching may occasionally degrade, overall scale-normalised error still improves. The improvement maps (Figure~\ref{fig:imp_baseline_maps}) visually confirm a strong and geographically widespread margin of superiority.

\medskip
\noindent\textbf{Improvement over STLM.}
Relative to the STLM benchmark, HSTM also delivers consistent gains, though of smaller magnitude, as expected from a stronger reference point. Across most districts, sMAPE decreases by 8--15\% and NRMSE by 12--25\%. Notably, \textit{Malda} and \textit{East Midnapore} show large reductions exceeding 20\% in NRMSE, underscoring the value of yearly-feature conditioning in regions prone to erratic monsoon variability. A few isolated cases---such as \textit{North Dinajpur}---show mild degradation in sMAPE despite improvement in NRMSE, suggesting that while overall magnitude tracking improves, phase/timing alignment remains challenging. These mixed cases highlight that absolute symmetry in error reduction across both metrics is not guaranteed, but the directional consistency of improvements remains strong overall.

Taken together, these results show that the hierarchical conditioning used in HSTM yields not only absolute performance gains over naïve baselines but also tangible and geographically coherent improvements over a competitive spatio-temporal benchmark. The improvement maps provide intuitive visual evidence of robustness across diverse rainfall regimes---from Himalayan foothills to Gangetic plains and coastal deltas---suggesting that the architecture generalises well across heterogeneous climatic zones.

\subsubsection{Yearly Forecast Consistency}

To assess temporal stability beyond monthly accuracy, we evaluate the models on 
yearly-aggregated forecasts. Figure~\ref{fig:hstm_model_act_f} compares observed and 
predicted annual rainfall for three representative districts --- Bankura (semi-arid inland), 
Jalpaiguri (high-rainfall northern belt), and Kolkata (coastal urban). Unlike the Na\"ive 
baseline, which exhibits persistent lag in capturing reversals, and the STLM model, which 
tends to overreact to interannual spikes, the HSTM forecasts demonstrate strong phase 
alignment with observed trajectories. Even when the absolute magnitude deviates slightly, 
the turning points (e.g., drought-to-recovery transitions) are correctly identified, indicating 
reliable directional forecasting.

A broader consistency check is provided in Figure~\ref{fig:hstm_model_ht}, which shows a 
district--year heatmap of symmetric MAPE (sMAPE) errors. In contrast to the Na\"ive and STLM 
models, which display intermittent ``failure years'' in volatile regions such as Malda, Hoogly etc., the HSTM maintains uniformly moderate error levels across both space and time. 
The absence of extended high-error streaks suggests resilience against error accumulation 
in long-horizon recursive forecasting.

Overall, these diagnostics confirm that HSTM offers not only superior mean accuracy but also 
greater temporal robustness. This stability is essential for downstream planning applications 
where consistency in interannual signals is often more valuable than marginal gains in 
pointwise precision.

\section{Conclusion}
This study presented a comprehensive comparison of three forecasting models for district-level rainfall prediction. While the baseline Na\"ive Seasonal model provided a simple reference point, and the STLM model incorporated local temporal dynamics effectively, their performance declined in several districts due to progressive error accumulation and limited spatial awareness. To address these limitations, the proposed HSTM framework introduced hierarchical learning that captures both temporal regularities and cross-district dependencies.

The empirical results demonstrated that HSTM consistently outperformed STLM across most districts in terms of sMAPE and NRMSE, with substantial improvement particularly in regions exhibiting high variability. Although a few districts showed marginal or negative gains, the overall performance trend establishes HSTM as the most reliable and generalizable model among the three. These findings reinforce the importance of incorporating spatial structure alongside temporal modeling in regional climatic forecasting.

\section{Future Work}
Several directions remain open for further enhancement of this forecasting framework. First, exogenous predictors such as large-scale climatic indices (ENSO, IOD), soil moisture, or land use variables may be integrated to enrich the model’s explanatory capacity, particularly for anomalous years. Second, the current hierarchical structure may be extended with probabilistic or Bayesian formulations to quantify predictive uncertainty more effectively. Third, recent advancements in graph neural networks and attention-based sequence models could be explored for finer-grained spatial dependency modeling.

Finally, a real-time operational deployment, coupled with a visualization dashboard or early-warning interface, would translate the proposed framework into a practical decision-support tool for agricultural planning, water resource management, and climate resilience policy.

\section*{Acknowledgment}
The authors sincerely thank the \textit{West Bengal Pollution Control Board, Government of West Bengal}, for providing the rainfall dataset used in this study. The dataset contains daily observed rainfall records from 1900 to 2019 collected from multiple observatories across various districts of West Bengal. This valuable data served as the foundation for the analyses and results presented in this report.

\section*{Appendix : District-wise Improvement Tables}

For completeness and reproducibility, this appendix provides the full numerical values corresponding to the relative performance improvements from naive seasonal to HSTM. While the main text presents only the spatial visualizations and aggregated summaries, the following tables list the exact percentage change in sMAPE and NRMSE for each district under different model comparisons.
\medskip
Specifically, we report:

\begin{itemize}
    \item \textbf{Table \ref{tab:stlm_naive}}: Performance gain of the STLM benchmark relative to the naïve seasonal baseline.
    \item \textbf{Table \ref{tab:hstm_naive}}: Performance gain of the proposed HSTM model relative to the naïve baseline.
    \item \textbf{Table \ref{tab:hstm_stlm}}: Incremental gain of HSTM over the stronger STLM benchmark.
\end{itemize}

Positive values indicate improvement (i.e., percentage reduction in error), while negative values denote deterioration. These tables provide fine-grained insight into district-level heterogeneity and support the conclusions drawn in the performance comparison section.

\begin{table}[H]
\centering
\caption{Percentage improvement of STLM over Naïve baseline (sMAPE and NRMSE).}
\label{tab:stlm_naive}
\begin{tabular}{lcc}
\hline
\textbf{District} & \textbf{sMAPE } & \textbf{NRMSE } \\
\hline
BANKURA & 16.87 & 11.40 \\
BIRBHUM & 8.34 & 9.54 \\
BURDWAN & 24.06 & 14.21 \\
COOCH BEHAR & 18.41 & 21.38 \\
DARJEELING & 25.28 & 28.31 \\
HOOGLY & 13.45 & 21.16 \\
HOWRAH & 19.06 & 17.21 \\
JALPAIGURI & 1.34 & 20.92 \\
MALDA & -4.58 & -3.41 \\
MANBHUM PURULIA & 8.87 & 7.94 \\
EAST MIDNAPORE & -0.87 & -3.70 \\
MURSHIDABAD & 22.07 & 19.43 \\
NADIA & 16.14 & 13.18 \\
24 PARGANAS N & 20.79 & 36.19 \\
24 PARGANAS S & 11.17 & -17.71 \\
DINAJPUR NORTH & 16.11 & 18.38 \\
DINAJPUR SOUTH & -25.14 & 13.12 \\
WEST MIDNAPORE & 23.13 & 10.02 \\
KOLKATA & 9.95 & -10.10 \\
\hline
\end{tabular}
\end{table}

\begin{table}[H]
\centering
\caption{Percentage improvement of HSTM over Naïve baseline (sMAPE and NRMSE).}
\label{tab:hstm_naive}
\begin{tabular}{lcc}
\hline
\textbf{District} & \textbf{sMAPE } & \textbf{NRMSE } \\
\hline
BANKURA & 28.64 & 22.86 \\
BIRBHUM & 18.91 & 31.99 \\
BURDWAN & 33.46 & 40.35 \\
COOCH BEHAR & 29.37 & 29.90 \\
DARJEELING & 30.45 & 38.74 \\
HOOGLY & 22.18 & 31.19 \\
HOWRAH & 23.15 & 28.87 \\
JALPAIGURI & 7.71 & 31.29 \\
MALDA & 20.01 & 30.16 \\
MANBHUM PURULIA & 16.67 & 22.41 \\
EAST MIDNAPORE & 15.90 & 23.20 \\
MURSHIDABAD & 26.99 & 34.17 \\
NADIA & 24.39 & 34.06 \\
24 PARGANAS N & 29.67 & 45.17 \\
24 PARGANAS S & 20.18 & 21.82 \\
DINAJPUR NORTH & 9.08 & 40.86 \\
DINAJPUR SOUTH & -6.13 & 18.55 \\
WEST MIDNAPORE & 31.63 & 30.38 \\
KOLKATA & 20.88 & 24.61 \\
\hline
\end{tabular}
\end{table}

\begin{table}[H]
\centering
\caption{Percentage improvement of HSTM over STLM (sMAPE and NRMSE).}
\label{tab:hstm_stlm}
\begin{tabular}{lcc}
\hline
\textbf{District} & \textbf{sMAPE} & \textbf{NRMSE } \\
\hline
BANKURA & 14.16 & 12.94 \\
BIRBHUM & 11.53 & 24.83 \\
BURDWAN & 12.38 & 30.47 \\
COOCH BEHAR & 13.43 & 10.83 \\
DARJEELING & 6.92 & 14.55 \\
HOOGLY & 10.09 & 12.71 \\
HOWRAH & 5.05 & 14.09 \\
JALPAIGURI & 6.45 & 13.11 \\
MALDA & 23.51 & 32.47 \\
MANBHUM PURULIA & 8.55 & 15.72 \\
EAST MIDNAPORE & 16.63 & 25.93 \\
MURSHIDABAD & 6.33 & 18.30 \\
NADIA & 9.84 & 24.05 \\
24 PARGANAS N & 11.21 & 14.08 \\
24 PARGANAS S & 10.14 & 33.58 \\
DINAJPUR NORTH & -8.38 & 27.54 \\
DINAJPUR SOUTH & 15.19 & 6.24 \\
WEST MIDNAPORE & 11.06 & 22.63 \\
KOLKATA & 12.14 & 31.52 \\
\hline
\end{tabular}
\end{table}

\newpage

\end{document}